\documentclass[useAMS,usenatbib]{mn2e}
\usepackage{graphicx}
\usepackage{amssymb}
\usepackage{amsmath}
\usepackage{subfigure}
\usepackage{ifpdf}
\usepackage{epstopdf}
\usepackage{longtable}
\usepackage{array}
\usepackage{hyperref}

\title[A dynamical study of CALIFA galaxies]
  {A dynamical view on stellar metallicity gradient diversity across the Hubble sequence with CALIFA}
\author[Y. Zhuang et al.]
{Yulong Zhuang$^{1}$\thanks{Email: zhuang@mpia.de}, Ryan Leaman$^{1}$, Glenn van de Ven$^{1,2}$, Stefano Zibetti$^{3}$,\
                                                                                                                                                                                                                                                                                                                                                                                                                                                                                                                                                                                                                                                                                                                                                                                                                                                                                                                                                                                                                                                                                                                                                                                                                                                                                                                                                                                                                                                                                                                                                                                                                                                                                                                                                                                                                                                                                                                                                                                                                                                                                                                                                                                                                                                                                                                                                                                                                                                                                                                                                                                                                                                                                                                                                                                                                                                                                                                                                                                                                                                                                                                                                                                                                                                                                                                                                                                                                                                                                                                                                                                                                                                                                                                                                                                                                                                                                                                                                                                                                                                                                                                                                                                                                                                                                                                                                                                                                                                                                                                                                                                                                                                                                                                                                                                                                                                                                                                                                                                                                                                                                                                                                                                                                                                                                                                                                                                                                                                                                                                                                                                                                                                                                                                                                                                                                                                                                                                                                                                                                                                                                                                                                                                                                                                                                                                                                                                                                                                                                                                                                                                                                                                                                                                                                                                                                                                                                                                                                                                                                                                                                                                                                                                                                                                                                                                                                                                                                                                                                                                                                                                                                                                                                                                                                                                                                                                                                                                                                                                                                                                                                                                                                                                                                                                                                                                                                                                                                                                                                                                                                                                                                                                                                                                                                                                                                                                                                                                                                                                                                                                                                                                                                                                                                                                                                                                                                                                                                                                                                                                                                                                                                                                                                                                                                                                                                                                                                                                                                                                                                                                                                                                                                                                                                                                                                                                                                                                                                                                                                                                                                                                                                                                                                                                                                                                                                                                                                                                                                                                                                                                                                                                                                                                                                                                                                                                                                                                                                                                                                                                                                                                                                                                                                                                                                                                                                                                                                                                                                                                                                                                                                                                                                                                                                                                                                                                                                                                                                                                                                                                                                                                                                                                                                                                                                                                                                                                                                                                                                                                                                                                                                                                                                                                                                                                                                                                                                                                                                                                                                                                                                                                                                                                                                                                                                                                                                                                                                                                                                                                                                                                                                                                                                                                                                                                                                                                                                                                                                                                                                                                                                                                                                                                                                                                                                                                                                                                                                                                                                                                                                                                                                                                                                                                                                                                                                                                                                                                                                                                                                                                                                                                                                                                                                                                                                                                                                                                                                                                                                                                                                                                                                                                                                                                                                 \newauthor Anna Gallazzi$^{3}$, Ling Zhu$^{1}$ , Jes\'{u}s Falc\'{o}n-Barroso$^{4,5}$ and Mariya Lyubenova$^{2}$\\
$^{1}$Max-Planck Institut f\"ur Astronomie, K\"onigstuhl 17, D-69117 Heidelberg, Germany\\
$^{2}$European Southern Observatory, Karl-Schwarzschild-Str. 2, 85748 Garching, Germany\\
$^{3}$INAF-Osservatorio Astrofisico di Arcetri, Largo Enrico Fermi 5, I-50125 Firenze, Italy\\
$^{4}$Instituto de Astrof\'{i}sica de Canarias, v\'{i}a L\`{a}ctea s/n, 38205 La Laguna, Tenerife, Spain\\
$^{5}$Departamento de Astrof\'{i}sica, Universidad de La Laguna, E-38200 La Laguna, Tenerife, Spain\\}
\date{Accepted 2015. Received 2015; in original form 2015 }

\pagerange{\pageref{firstpage}--\pageref{lastpage}} \pubyear{2014}

\begin{document}

\label{firstpage}

\maketitle

\begin{abstract}

We analyze radial stellar metallicity and kinematic profiles out to 1$R_{\mathrm{e}}$ in 244 CALIFA galaxies ranging from morphological type E to Sd, to study the evolutionary mechanisms of stellar population gradients. We find that linear metallicity gradients exhibit a clear correlation with galaxy morphological type - with early type galaxies showing the steepest gradients. We show that the metallicity gradients simply reflect the local mass-metallicity relation within a galaxy. This suggests that the radial stellar population distribution within a galaxy's effective radius is primarily a result of the \emph{in-situ} local star formation history.  In this simple picture, the dynamically derived stellar surface mass density gradient directly predicts the metallicity gradient of a galaxy. We show that this correlation and its scatter can be reproduced entirely by using independent empirical galaxy structural and chemical scaling relations. 
Using Schwarzschild dynamical models, we also explore the link between a galaxy's local stellar populations and their orbital structures. 
We find that galaxies' angular momentum and metallicity gradients show no obvious causal link. This suggests that metallicity gradients in the inner disk are not strongly shaped by radial migration, which is confirmed by the lack of correlation between the metallicity gradients and observable probes of radial migration in the galaxies, such as bars and spiral arms.
Finally, we find that galaxies with positive metallicity gradients become increasingly common towards low mass and late morphological types - consistent with stellar feedback more efficiently modifying the baryon cycle in the central regions of these galaxies.

\end{abstract}


\section{Introduction}

A galaxy's radial stellar metallicity distribution is an important tracer of its evolutionary history, as the long-lived stars preserve a record of how efficiently heavy elements were produced, disseminated out to the ISM via stellar feedback, and locally enriched the gas which formed subsequent generations.  The present day spatial distribution of stars of different metallicities and ages can therefore provide clues to the key processes governing the dynamical evolution of galaxies.
 

As stars from existing populations can also be relocated due to orbital modifications~\citep[e.g., disk heating or radial migration by bars or spiral arms; c.f.,][]{1987gady.book.....B} a connection between the spatial distribution of different chemical sub-populations and their stellar dynamics is expected. Joint analysis of the galaxy's chemo-dynamical properties can thus provide stronger insight into how the galaxy was assembled, than from just kinematics or chemistry alone.
 
The efficiency of radial migration in galaxies of different morphological types has been difficult to quantify in external galaxies.  For example, \cite{2014A&A...570A...6S} presented the stellar metallicity and age distributions in a sample of 62 nearly face-on, spiral galaxies with and without bars, using data from the CALIFA survey. They found no difference in the metallicity or age gradients between galaxies with and without bars. Stronger evidence for stellar migratory processes comes primarily from simulations and detailed studies of the Milky Way (MW) ~\citep[e.g.,][]{2002MNRAS.336..785S,2012ApJ...758...41R}. In both cases, non-axisymmetric substructures (e.g., bar and spiral arms) seem to efficiently redistribute angular momentum of the galactic stellar populations ~\citep[e.g.,][]{2007ApJ.666..189B}. The combined effect of these overdensities can cause a radial displacement of stars ~\citep[e.g.,][]{2008ApJ...675L..65R,2010ApJ...722..112M,2011A&A...527A.147M,2012A&A...548A.126M}.  These same non-axisymmetric features can also directly relocate metals in the ISM via streaming motions \citep[e.g.][]{2014ASPC..480..211C}. 

Recently ~\cite{2016ApJ...820..131E} also examined the effects of stellar feedback and bursty star formation on low-mass galaxies using the FIRE (Feedback in Realistic Environments) simulations. They found that for low mass galaxies, repeated stellar feedback can rapidly remove gas from the galaxy's center, causing an expansion of the stellar and dark matter components - thus in addition to driving metal-enriched outflows \citep[e.g.,][]{2005MNRAS.363....2K,2007AAS...21111102B,2012A&A...540A..56P,2013A&A...554A..47G}, feedback can also indirectly modify radial stellar population gradients.


Externally, galaxy interactions, accretion events and mergers can not only strongly change the gravitational potential of the galaxies, but also rapidly redistribute the location and angular momentum of stars. A galaxy's merger history is therefore considered to play a significant role in determining the shape of the metallicity gradient in the galactic halo ~\citep[e.g.,][]{2014MNRAS.444.2938H}, while simultaneously altering the angular momentum of its stars. For example, simulations have shown that a major merger could efficiently flatten the metallicity gradient~\citep{2010ApJ...710L.156R,2013MNRAS.436.3507N}, while continuous minor mergers will steepen the gradient and shape the outskirts of the halo as the low mass satellites are accreted primarily at large radii~\citep[see also,][]{2013MNRAS.434.3348C,2013MNRAS.428..641O,2014MNRAS.444..237P}.  Similarly, \cite{2004MNRAS.347..740K} studied the formation and chemodynamical evolution of 124 elliptical galaxies with a GRAPE-SPH simulation code and found that galaxies that form monolithically tend to have steeper gradients than the ones that undergo major mergers.

While simulations have continued to show the utility of metallicity gradients in understanding the mass growth history of galaxies, less work has been devoted to how the metallicity gradients and angular momentum of the \textit{in-situ} stellar component co-evolve in the inner regions ($\leq R_{e}$) of galaxy's across the Hubble sequence.



Identifying the relative strength of \textit{in-situ} star formation (SF), radial stellar migration and accretion events therefore necessitates studying stellar population gradients in morphologically diverse samples of galaxies. Not surprisingly, recent surveys which observed a large variety of galaxies have found an equally large diversity of observational trends. For example, ~\cite{2017MNRAS.465.4572Z} studied the stellar population distributions and the environment of 1105 MaNGA galaxies. They found that both the age and metallicity gradients show weak or no correlation with either the large-scale structure type or local environmental density.  ~\cite{2017MNRAS.465..688G} studied the radial stellar population gradients of 721 MaNGA galaxies with their results suggesting, higher mass late type galaxies show steeper metallicity gradients.  Similarly, Martin-Navarro et. al. (2017) studied the normalization and the slope of the stellar populations gradients in early type CALIFA galaxies, finding a gradual steepening of the metallicity gradients with increasing galaxy velocity dispersion.

In order to reconcile these diverse observational results, and further clarify the relative strength of different mechanisms in setting stellar metallicity gradients, in this paper we explore a new formalism for understanding stellar metallicity gradients, using a homogeneous analysis of the \textit{chemo-dynamic} signatures present in the inner regions of CALIFA galaxies.

As all theoretical mechanisms presented should simultaneously alter the dynamical and chemical properties of the galaxy, we propose to leverage information on the orbital properties of stars, in the inner regions of galaxies spanning a range of masses and morphological types.  This allows us to study the impact of secular processes which may impart signatures on the angular momentum distribution of stars of different metallicities. Specifically, as a new analysis technique, we incorporate our group's orbit based Schwarzschild dynamical models to understand and quantify any link between metallicity profiles, migratory signatures and the dynamical structure of the galaxy.

Section 2 and 3 describe the extraction of the metallicity and kinematic profiles from 244 nearby CALIFA galaxies, which span a large range in mass (10$^{9}$M$_{\sun}$ to 10$^{12}$M$_{\sun}$),  and morphological type (from E to Sd). The subsequent sections present a simple picture to illustrate the dominant in-situ formation channel for stellar population gradients in our sample, which can naturally recover the observed diversity.

\begin{figure*}
\begin{center}
\includegraphics[width=0.99\textwidth, height=0.4\textheight]{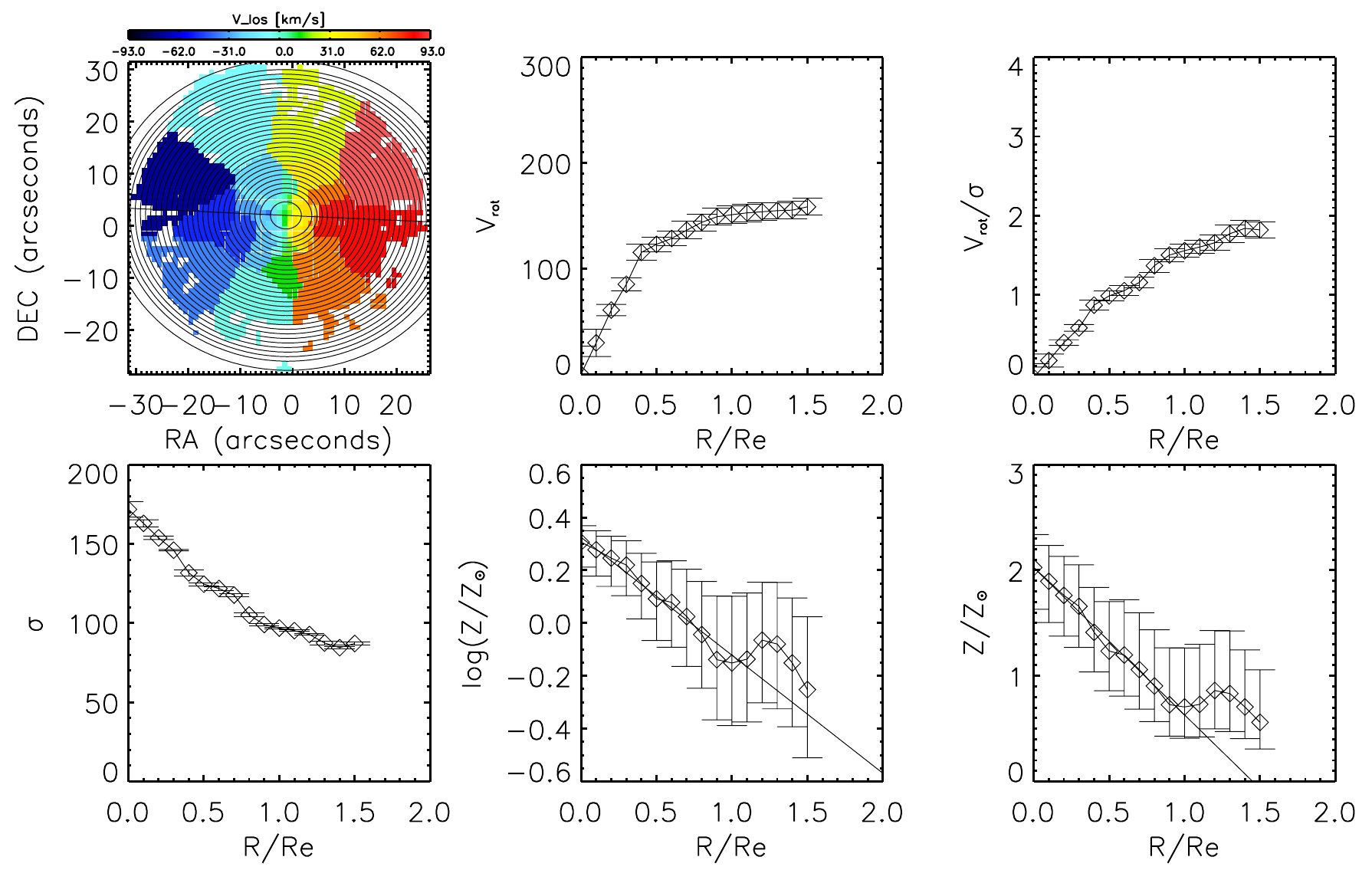}
\caption{An example of the  $V_{los}$ map based kinematic ellipse system (top left, with color-coding by $V_{los}$),  and derived rotation velocity (top middle), ratio between rotation velocity and velocity dispersion (top right), velocity dispersion (bottom left), logarithmic stellar metallicity (bottom middle) and linear metallicity (bottom right) profiles for NGC0932.}
\label{fig:all_NGC0932}
\end{center}
\end{figure*}

\section{Observations and data analysis}

The Calar Alto Legacy Integral Field Area Survey ~\citep[CALIFA;][]{2012A&A...538A...8S,2013A&A...549A..87H,2014A&A...569A...1W,2016A&A...594A..36S} uses the PMAS/PPAK integral field spectrophotometer to provide the largest and most comprehensive wide-field IFU survey of galaxies. It covers two overlapping spectral regions, one in the red (3745 - 7300 $\AÅ$) with resolution of R = 850 (V500 setup) and one in the blue (3400 - 4750 $\AÅ$) with R = 1650 (V1200 setup). Due to the quality and completeness of the data, our final sample includes 244 CALIFA galaxies from E to Sd morphological types, covering stellar mass range ($10^{9}$M$_{\sun}<$M$_{*}<10^{12}$M$_{\sun}$) ~\citep[for detailed information see Table 1 and][]{2014A&A...569A...1W}. We make use of the total stellar masses inferred from SED fitting to the photometry from UV to IR.

In a previous work, ~\cite{2017A&A...597A..48F} extracted the stellar kinematics for every galaxy in a uniform way using both instrumental setups, i.e., V500 and V1200. For this analysis we utilize the derived stellar velocity $V_{los}$, velocity dispersion $\sigma$ and their error maps for each galaxy.

Zibetti et al. (in preparation) derive maps of light-weighted stellar metallicity $Z$ as well as of other stellar-population properties from the joint analysis of the CALIFA spectroscopic datacubes and the SDSS 5-band images of the full final data release of CALIFA \citep[DR3][]{2016A&A...594A..36S}. In \cite{2017MNRAS.468.1902Z} the method is described in detail regarding the derivation of the maps of light- and mass-weighted ages, but it applies in the same way to the derivation of stellar metallicities, modulo substituting the property to be derived. We therefore refer the reader to \cite{2017MNRAS.468.1902Z} and summarize here the main points of the analysis and related uncertainties.

For each spaxel in each galaxy we compute the posterior probability function (PDF) of the light-weighted stellar metallicity $Z\equiv\frac { \sum_j  L_{ j }Z_{ j } }{ \sum_j  L_{ j } }$, where the $j$ index identifies the different Simple Stellar Population (SSP) components, of metallicity $Z_j$ and $r$-band luminosity $L_j$, respectively. We characterize the PDF by its median and $16^{th}$ and $84^{th}$ percentiles (i.e. $\pm 1\, \sigma$~for a gaussian distribution). We use a large and comprehensive library of 500,000 spectral models to infer $Z$ based on the comparison between \emph{measured} stellar absorption features and photometric fluxes in SDSS $ugriz$ bands and the corresponding \emph{synthetic} quantities measured on the model spectra.  The posterior probability function for $Z$ given the data is derived by folding the prior distribution of models in $Z$ with the likelihood of the data given each model $i$, 
$\mathcal{L}_i\propto\exp(-\chi_i^2/2)$. In practice, the probability $dP(Z)$ for a metallicity range $[Z - Z+dZ]$ given the data is derived from the sum of the likelihoods of the data over all models $i$ with a metallicity $Z_i$ in the given range.

The set of spectral indices used to compute the likelihood is the one introduced by \cite{2005MNRAS.362...41G}, namely three (mostly) age-sensitive indices ($\mathrm{D4000_n}$, $\mathrm{H\beta}$, and $\mathrm{H\delta_A}+\mathrm{H\gamma_A}$) and two (mostly) metal-sensitive composite indices that show minimal dependence on $\alpha$-element abundance relative to iron-peak elements ($[\mathrm{Mg_2Fe}]$ and $[\mathrm{MgFe}]^\prime$).

The spectral library is built based on the SSP spectra provided in the latest revision of the \cite[BC03]{2003MNRAS.344.1000B} stellar population synthesis models adopting the MILES spectral libraries \citep{2006A&A...457..809S,2011A&A...532A..95F} with a \cite{2003ApJ...586L.133C} initial mass function. As for the star-formation histories (SFHs) we adopt a \cite{1986A&A...161...89S} model with random bursts:

\begin{equation} 
\begin{split}
&SFR(t)=\frac{(t-t_0)}{\tau}\exp\left(-\frac{(t-t_0)^2}{\tau^2}\right)\\
&+ \sum_{j=1}^{N_\mathrm{burst}}A_j\,\delta_\mathrm{Dirac}((t-t_0)-t_{\mathrm{burst},j}), (t>t_0)
\end{split} 
\end{equation}

where $\tau$ is the time of the maximum SFR in the continuous component, and $A_j$ and $t_{\mathrm{burst},j}$ are the integrated intensity (relative to the integral of the continuous component) and the time of the instantaneous bursts. All these parameters as well as the number of bursts $N_\mathrm{burst}\leq 6$ are randomly generated.

Concerning the metallicity, we implement a simple recipe to take its evolution along the SFH into account. Following equation 11 of \cite{2008ApJ...674..151E} we impose a steady growth of metallicity according to:

\begin{equation}
\begin{split}
Z(t)&=Z\left(M(t)\right)\\
&=Z_{\mathrm{final}}-\left(Z_{\mathrm{final}}-Z_{\mathrm{0}}\right)\left(1-\frac{M(t)}{M_\mathrm{final}}\right)^\alpha,  \alpha \geq 0
\end{split}
\end{equation}

where $M(t)$ and $M_\mathrm{final}$ denote the stellar mass formed by time $t$ and by the final time of observation, $Z_0$ and $Z_\mathrm{final}$ are the initial and final values of stellar metallicity and $\alpha$ is the "swiftness"  parameter of the enrichment. $Z_{*\,\mathrm{0}}$ and $Z_{*\,\mathrm{final}}$ are chosen from a logarithmically uniform distribution in the interval $1/50$--$2.5~Z_\odot$. 

We also implement a dust treatment based on the \cite{2000ApJ...539..718C} 2-component interstellar medium \citep[diffuse ISM $+$ birth-cloud, see also][]{2008MNRAS.388.1595D}.

We adopt a broad range for all random-generated parameters in order to cover a range of SFHs and observable parameters as comprehensive as possible. In fact, at basically each spaxel we find a best-fit model with reduced $\chi^2 \leq 1$, indicating that our models can reproduce the observations very closely. 
The input distribution of model-generating parameters translates into non-uniform distributions in the physical parameter space, which however is roughly uniform in \emph{r-band-light-weighted} $\log$~age between $\sim5 \times 10^8$~and $\sim1.5 \times 10^{10}$~years and covers the full range of \emph{r-band-light-weighted} metallicities between  $1/50$ and $2.5~Z_\odot$.

Typical errors (i.e. $16^{th}-84^{th}$ percentile half-range) in $\log Z$ are around 0.15 dex in individual spaxels. From Montecarlo simulations which assume the same typical properties of the CALIFA+SDSS dataset we obtain a similar scatter between input and retrieved $\log Z$ (i.e. median value of the PDF) of $\sim 0.15$~dex. We also quantify possible biases across the age-metallicity plane, which are typically within $\pm 0.05$~dex. Only at the highest metallicities ($Z>2 Z_\odot$) there is a clear tendency to underestimate the true metallicity by up to $0.10-0.15$~dex, as a consequence of the PDFs being truncated at the high-end of the grid. It must be noted that a significant component of the scatter, between 0.05 and 0.1 dex, is not related to random measurement errors but is systematic and due to the intrinsic degeneracy between the observables and the physical parameters.

To explore the chemodynamical correlations in the galaxies we make use of recent dynamical models of our CALIFA sample.
\cite{2018MNRAS.473.3000Z} constructed triaxial Schwarzschild models \citep[c.f.,][]{2008MNRAS.385..647V} of 300 CALIFA galaxies in a uniform way, which simultaneously fit the observed surface brightness and stellar kinematics from the CALIFA survey. The procedure obtains the weights of different orbits that contribute to the best-fitting model. They use the time-averaged radius to represent the size of each orbit; and describe the angular momentum of an orbit by its circularity $\lambda_z$, which is defined as the angular momentum around the short axis of the galaxy normalized by the maximum that is allowed by a circular orbit with the same binding energy. The circularity distribution within 1R$_e$ for these 244 galaxies are shown in \cite{2017arXiv171106728Z,2018NatAs...2..233Z}. With the best-fitting stellar mass-to-light ratio we can not only estimate a dynamically derived stellar surface mass density, but also the stellar surface mass density maps of each orbital component; the cold, warm and hot orbital components we use here are separated by $\lambda_z > 0.8$, $0.25<\lambda_z<0.8$ and $|\lambda_z| < 0.25$, respectively.

\begin{figure*}
\begin{center}
\includegraphics[width=1\textwidth, height=0.24\textheight]{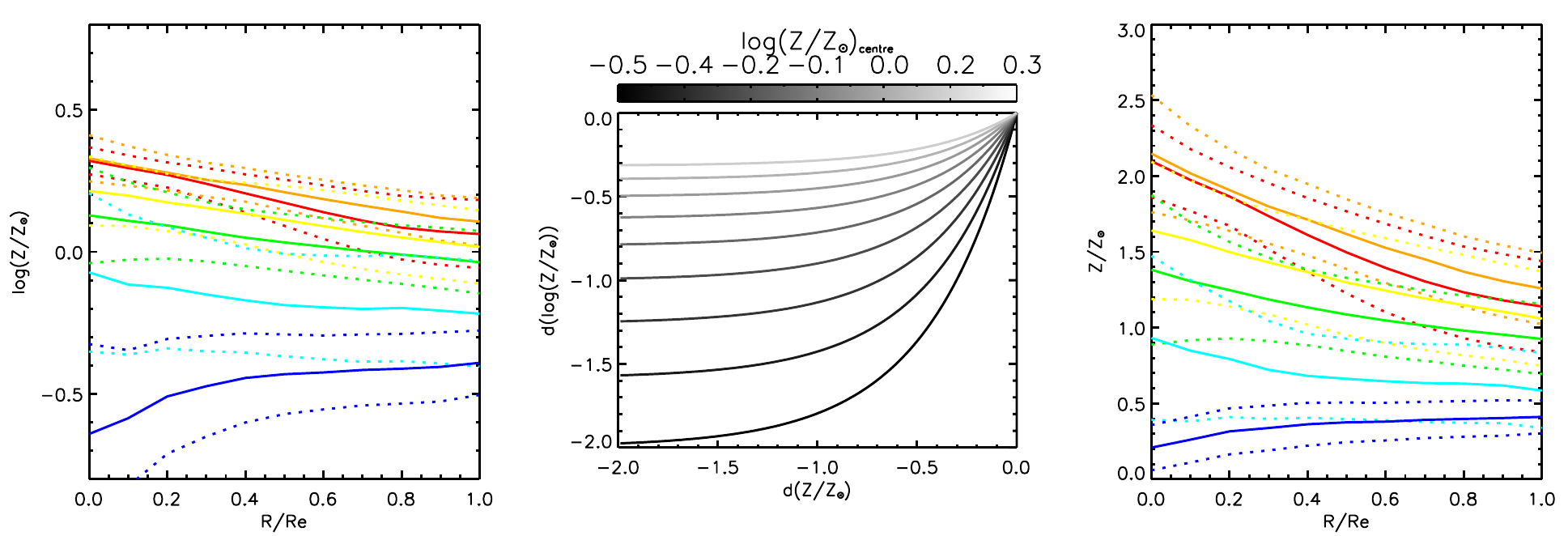}
\caption{Average log(Z) (left) and Z (right) profiles of each galaxy morphological type in our sample (solid lines), and standard deviations (dashed lines). Profiles are color-coded by galaxy morphological types: E (Red), S0 (Orange), Sa (Yellow), Sb (Green), Sc (Cyan) and Sd (Blue). The middle panel is the relation between log(Z) gradient and Z gradient for galaxies of different central metallicities. High mass galaxies, with high central metallicity, will show a smaller log(Z) gradients, conversely lower mass galaxies with low central metallicities will show a larger log(Z) gradients at fixed linear metallicity gradient, solely due to the mathematical logarithm transform.}
\label{fig:Ave_Zslope}
\end{center}
\end{figure*}

\section{Stellar Kinematic Profiles and Metallicity Gradients}

We use the tilted-ring method ~\citep{1974ApJ...193..309R,1976ApJ...204..703R} to compute radial profiles of a galaxy based on the galaxy's kinematic structure. 
An axisymmetric galactic dynamic model assumes that besides the system velocity, the disk velocity has three components: the rotation velocity $V_{rot}$, radial velocity $V_{rad}$ and vertical velocity $V_{z}$. In this model, a galaxy's line of sight velocity $V_{los}$ can be written as:  
\begin{equation}
\begin{split}
V_{los}(R,\psi,i) = V_{sys}&+V_{rot}(R)\mathrm{cos}{\psi}\mathrm{sin}i\\
 &+V_{rad}(R)\mathrm{sin}{\psi}\mathrm{sin}i\\
  &+V_{z}(R)\mathrm{cos}i
\end{split}
\end{equation}
For each point in the polar coordinates system ($R$, $\psi$), the three velocity components are projected by a galaxy's inclination angle \textit{i}.

Using the simplest version of the axisymmetric model with $V_{rad}(R)$ and $V_{z}(R)$ equal to zero in Eq.(1), we fit the global galaxy parameters (center, PA, inclination) from the $V_{los}$ map. With the best-fit parameters, we can divide the galaxy image into a series of concentric ellipses (see Figure ~\ref{fig:all_NGC0932}). Then we apply this ellipse system to other maps including velocity dispersion, log(Z) (and their 84\% and 16\% confidence maps log(Z)$_{16\%}$ and log(Z)$ _{84\%}$) to get their radial profiles. We construct the average value of a quantity at elliptical radius R by averaging the spaxels in that annulus. For logarithmic metallicities (log(Z) as well as log(Z)$_{16\%}$ and log(Z)$ _{84\%}$), we first calculate the linear flux-weighted mean values Z, Z$_{16\%}$, Z$_{84\%}$ from the spaxels within an annulus, then take the log of that value. We do this to ensure that we accurately trace the systematic errors due to e.g. the age-metallicity degeneracy.

We calculate the mean stellar velocity dispersion ($\sigma$) value from spaxels within each elliptical ring.
The extraction of the rotation velocity ($V_{rot}$) profile is slightly more complex, because of the existence of non-axisymmetric features (like bars, spiral arms) in the observed data. In order to cope with these detailed morphological structures and perturbations, a harmonic decomposition (HD) method is used in this work, in which we expand the $V_{los}$ into a Fourier series:

\begin{equation}
 V_{los}=V_{sys}+\sum_{n=1}^k [c_{n}(R)\mathrm{cos}n\psi+s_{n}(R)\mathrm{sin}n\psi]\mathrm{sin}i
\end{equation}
 
Here the $c_{1}$ term represents $V_{rot}$ while $s_{1}$ represents $V_{rad}$. Compared with an axisymmetric model, this method is better in fitting galaxies with non-axisymmetric features ~\citep{1989SvA....33..476S,1993ApJ...406..457C,1997MNRAS.292..349S,2003SSRv..105....1F,2005MNRAS.357.1113K}, which is important when studying the impact of migratory processes.
In this work, we fit the $V_{rot}$ at each elliptical annulus by using the HD method from~\cite{2005MNRAS.364..773F} .

As the kinematic maps ($V_{los}$, $\sigma$) were Voronoi binned into several superspaxels to increase S/N (see the top left panel of Figure~\ref{fig:all_NGC0932}), for each galaxy, we computed the ellipse system 200 times, each time giving a Gaussian distributed perturbation to the $V_{los}$ map, with standard deviation taken from the $V_{los}$ error map. Then for that ellipse system, we derive all the profiles as described above. The final profile and uncertainties due to stellar population parameters, are taken as the median of the 200 Z, Z$_{16\%}$ and  Z$_{84\%}$ profiles. While, the kinematic profiles and uncertainties are taken from the average and standard deviations of the 200 $V_{rot}$and $\sigma$ profiles.

Our method traces the profiles based on the rotation plane, which may produce differences compared to gradients extracted along the luminosity major axis, if the two planes are not overlapping each other. However, as we are not considering interacting galaxies, and ~\citep{2015A&A...582A..21B} has shown that around 90\% of non-interacting CALIFA galaxies show differences between their photometric and kinematic position angles of less than 22 degrees, here all galaxies show well defined angular momentum vectors which will be quite similar to the photometric position angles.  This is backed up by our Schwarzschild dynamical models (Section 2).

The profiles are derived through the geometry shape of the galaxies' (ellipse systems), which are determined by galaxy's kinematic map (rotation). This means our method would not apply to galaxies with no clear rotation signature. The E-type galaxies contained in our sample still contain significant rotational components to permit this. 
Some galaxies may show different rotation planes at different radii, which could potentially increase the uncertainty. However again, the photometric and kinematic position angle alignment found throughout non-interacting galaxies in CALIFA, suggests the ellipse system will not be significantly different using our method compared to photometrically defined ellipse systems.

\begin{figure}
\begin{center}
\includegraphics[width=0.4\textwidth, height=0.5\textheight]{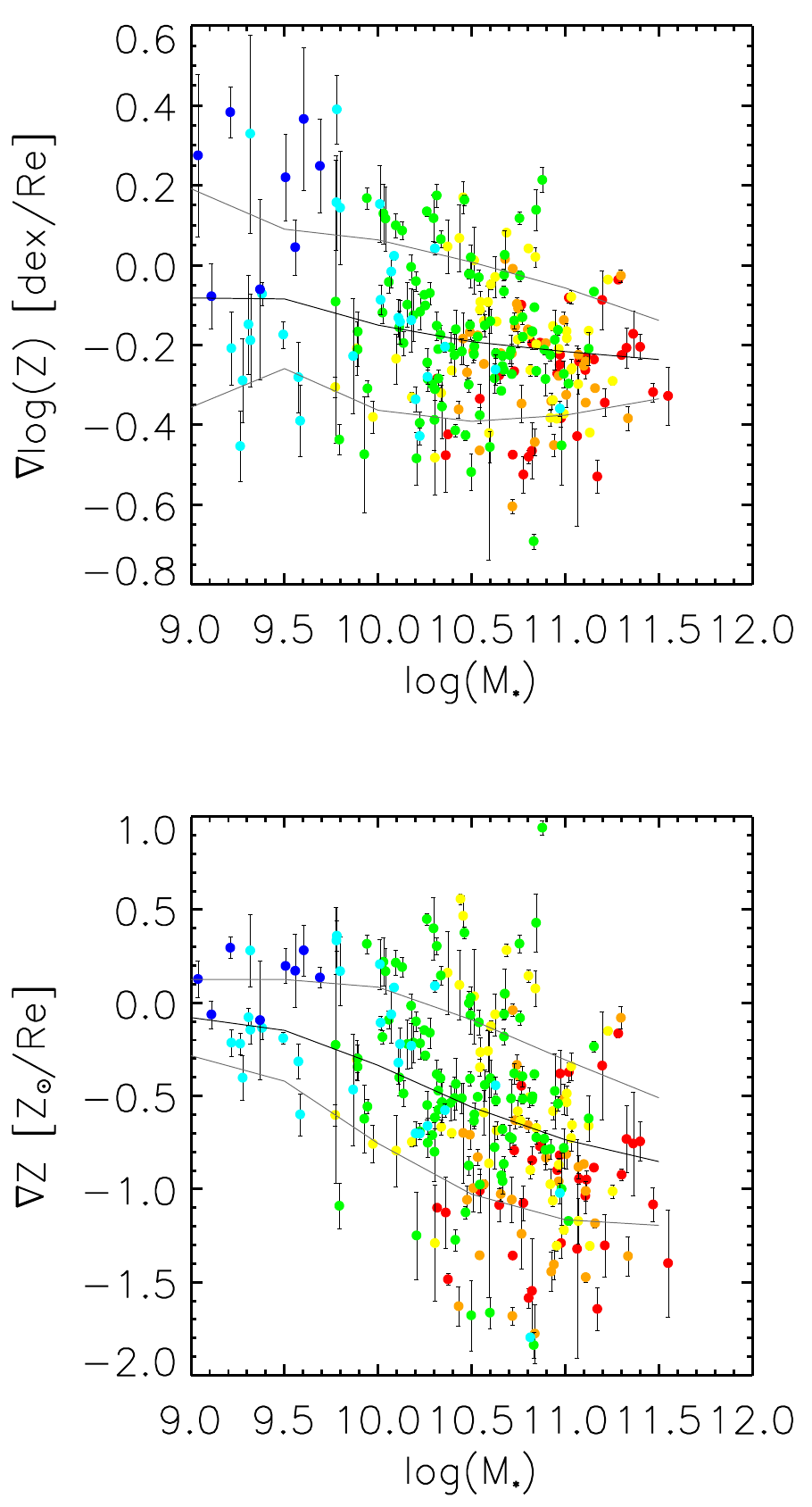}
\caption{The log(Z) and Z gradients against galaxy stellar mass, color-coded by galaxy morphological type as in Figure~\ref{fig:Ave_Zslope}. Black and grey lines are their mean and $1-\sigma$ intrinsic scatter at each mass range. }
\label{fig:FeH_Z_DB}
\end{center}
\end{figure}    

\subsection{Fitting metallicity gradients}

To fit the metallicity gradients ($\nabla$Z), for each of the 200 trials, we linearly fit the Z profile within certain radial ranges (1$R_{\mathrm{e}}$) in units of $R_{\mathrm{e}}$ with the weighted linear least squares routine MPFIT in IDL ~\citep{2009ASPC..411..251M}.  The weights used in fitting the linear gradient, came from the Z$_{16\%}$ and Z$_{84\%}$ uncertainty profiles in each trial. The final value of the gradient ($\nabla$Z), and uncertainty comes from the median and standard deviation of the 200 fits. Thus, the final uncertainty not only takes into account the uncertainties of the metallicity maps but also the uncertainty of the geometrical parameters of the galaxy.

We fit the metallicity gradients from both linear Z profiles and log(Z) profiles. In this work, we focus on the gradients computed from the linear Z profiles ($\nabla$Z). These show the difference in metal fraction from the inside to outside of the galaxy - as opposed to gradients of log(Z), which represent the ratio of metal fractions.  While neither is objectively superior, the former may show a more intuitive visual accounting of the absolute change in metal fraction across galaxy disks when comparing galaxies of a range of masses and absolute metallicities.  For example, the value of a log(Z) gradient for a high metallicity, massive galaxy, may be numerically smaller than that of a low metallicity, low mass galaxy - even if the radial difference in Z (linear gradient) is the same.  However the primary reason we use the linear Z gradients in this work, is to test the correlations between local metal fraction and local density and kinematics - which is not as easily accomplished with ratios. To conceptualize the differences and aid comparison to previous works, we show comparison of the \textit{profiles} in log(Z) and linear(Z) in Figure 2, with the middle panel showing the numerical equivalence between \textit{gradients} measured on each profile. In addition, given the range of physical galaxy sizes in our sample, we compute the gradients with respect to their relative effective radius, $R_{\mathrm{e}}$.

We note that,  all our galaxies have profiles extending to 1$R_{\mathrm{e}}$ and some of them even beyond 1.5$R_{\mathrm{e}}$. In this paper, unless stated otherwise, we fit the gradient within 1$R_{\mathrm{e}}$. In certain figures, we also linearly fit the profile between 0.5 to 1.5$R_{\mathrm{e}}$ and label these cases explicitly.

\section{Results}
 
The left panel of Figure~\ref{fig:Ave_Zslope} shows the average log(Z) profiles for galaxies of different morphological types in our sample. In general, early type galaxies (red) are more metal-rich than late type galaxies (blue), due to the mass-metallicity relation and mass-morphology relations ~\citep[e.g.,][]{2014ApJ...791L..16G}. It is clear that late type galaxies tend to have a larger diversity in profiles than earlier type galaxies - in agreement with results from gas-phase metallicity gradient studies ~\citep{2016MNRAS.456.2982T}.
The right panel of Figure~\ref{fig:Ave_Zslope} shows the average profiles for different morphological types in linear Z. Compared to the log(Z) profiles, the early type galaxies tend to have steeper Z profiles, and the linear profiles don't show strong decrease in diversity for early type galaxies.

We plot the gradients in both logarithmic and linear metallicity against galaxy stellar mass in Figure~\ref{fig:FeH_Z_DB}. Though we are mostly focusing on galaxies' linear metallicity gradient in this work, we still calculate the log(Z) gradients (top panel) to show the comparison with other works. The log(Z) gradients show a relatively weak correlation with mass, and an intrinsic scatter which decreases for massive/earlier type galaxies. This is in agreement with previous works on log(Z) gradients in morphologically diverse surveys \citep{2015A&A...581A.103G,2017MNRAS.465.4572Z}.  Also the correlation between stellar mass and metallicity gradients for late-type galaxies agrees with studies in MANGA galaxies ~\citep{2017MNRAS.465..688G}.

The bottom panel of Figure 3 plots the corresponding linear metallicity Z gradient against stellar mass, which shows a steeper anti-correlation with galaxy mass, and nearly constant intrinsic scatter in gradient value. The difference between the two plots is purely due to the logarithimic effect.

\begin{figure}
\begin{center}
\includegraphics[width=0.4\textwidth, height=0.5\textheight]{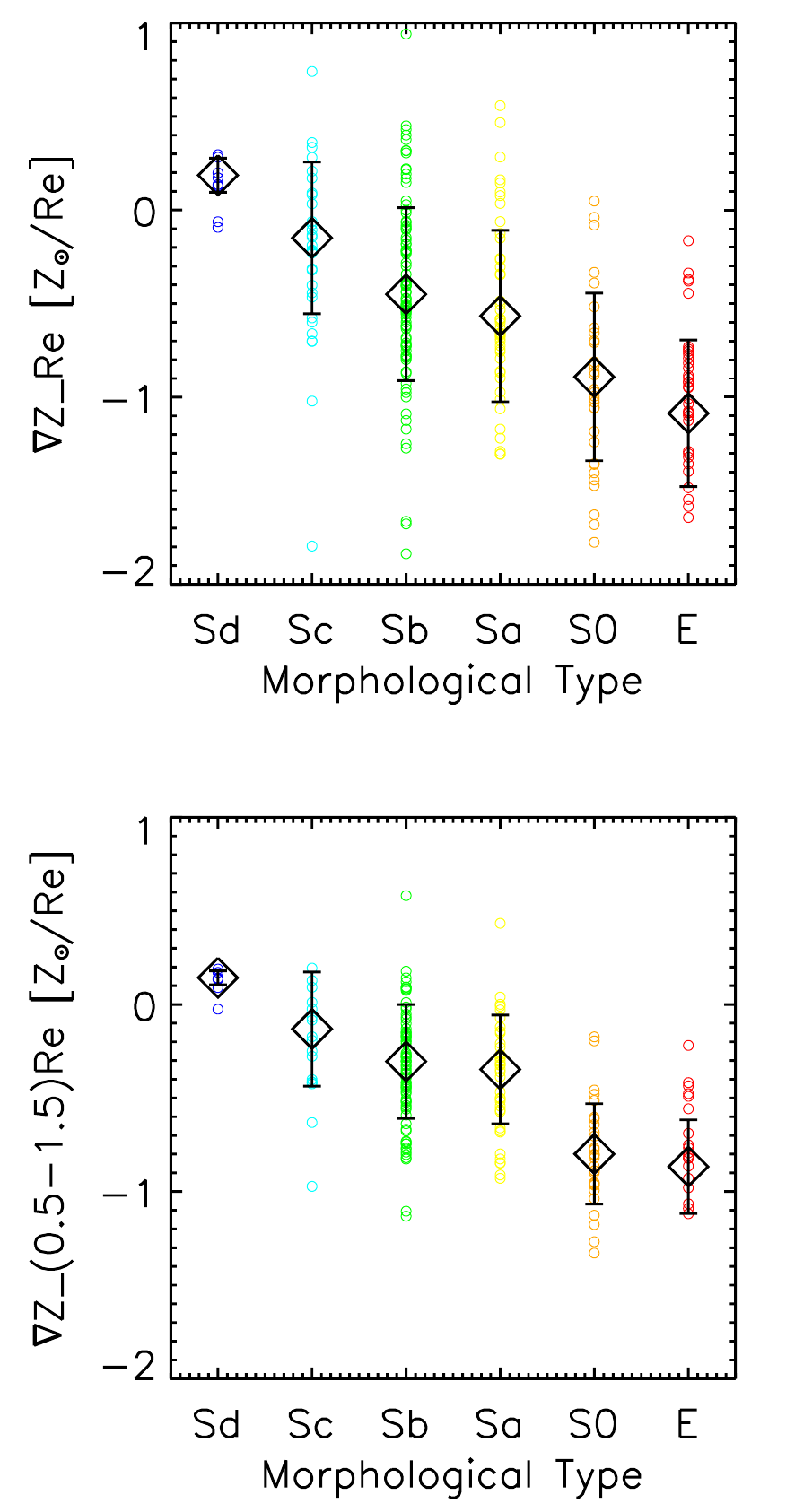}
\caption{Radial stellar metallicity gradient $\nabla$Z  within 1 $R_{\mathrm{e}}$ (top) and (0.5-1.5) $R_{\mathrm{e}}$ (bottom) against different morphological types. Error bars show the average and standard deviation after the volume correction of the CALIFA sample.  }
\label{fig:SP_G}
\end{center}
\end{figure}

The top panel of Figure~\ref{fig:SP_G} shows the Z gradients measured within 1 $R_{\mathrm{e}}$, as a function of morphological type. We find a clear correlation between metallicity gradients and galaxy morphological type: the Z gradient becomes steeper as we move towards early type galaxies. By comparing to Figure~\ref{fig:FeH_Z_DB}, we see that galaxies with different Z gradients are more segregated by morphology than stellar mass.

As in the very central part of galaxies there may exist different structures (Bar, bulge ...) for different morphological types, which may consist of quite different stellar populations (especially in terms of age), this may influence the correlation between Z gradients and morphological types.
To rule out this possibility, we also calculate the Z gradients from 0.5 to 1.5 $R_{\mathrm{e}}$ (the bottom panel of Figure~\ref{fig:SP_G}) to test the dependence of this trend on the central substructures. When the central regions are removed, the gradients on average become flatter, though the trend between Z and morphological type still remains equally strong. This suggests the correlation between the Z gradients and morphological type is more likely a general trend with mass concentration than a result of substructures of different formation epochs.

\section{Discussion}
 
\subsection{Understanding the morphology and Z gradient correlations}  
 
The sample of CALIFA galaxies analyzed here shows a strong correlation between linear metallicity gradient  ($\nabla$Z) and galaxy morphological type. The most obvious differences between different morphological types should be their average stellar metallicities (due to the morphology-mass and mass-metallicity relations, see also Figure~\ref{fig:Ave_Zslope}), their mass/light concentrations ~\citep[e.g.,][]{2014ApJ...781...12H} and their dynamical properties \citep{2017arXiv171106728Z}.

\begin{figure*}
\begin{center}
\includegraphics[width=1.0\textwidth, height=0.3\textheight]{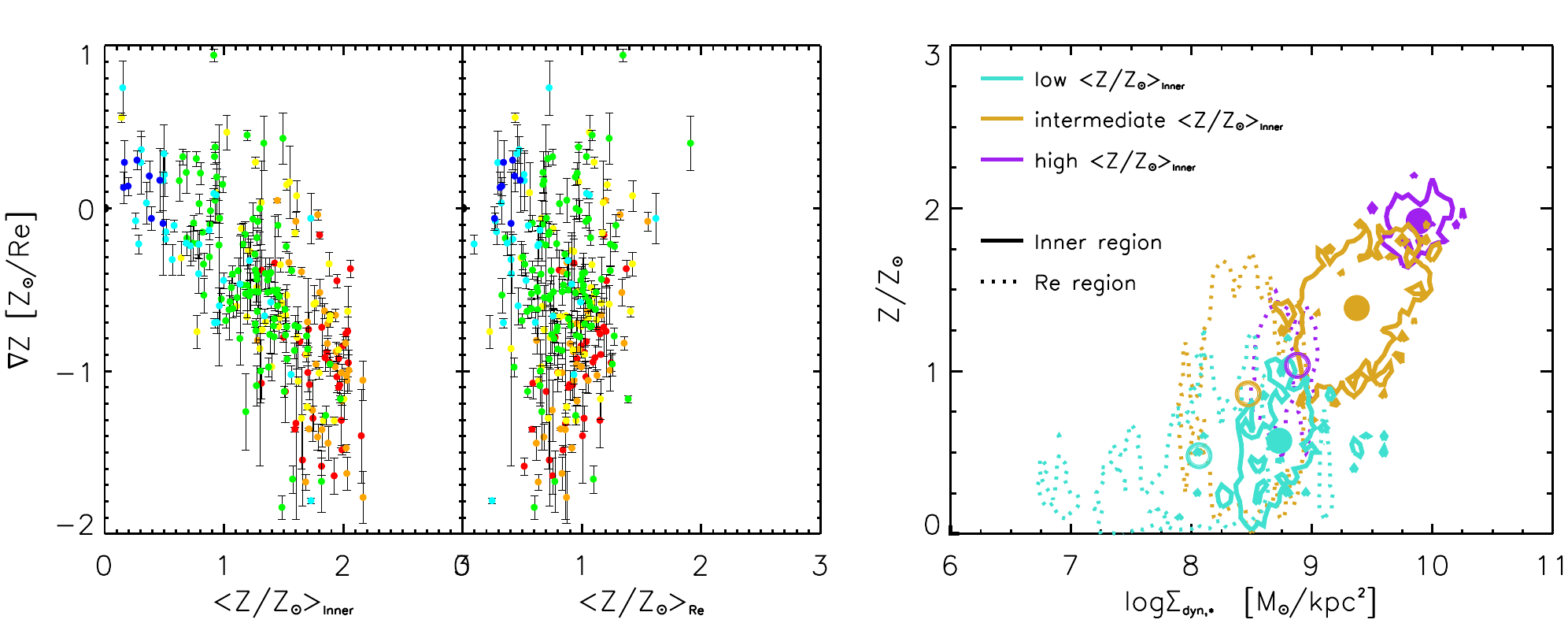}
\caption{\textit{Left}: Metallicity Z gradient against average stellar metallicity $<\mathrm{Z/Z}_{\sun}>$ for galaxies in an inner region within 0.2 $R_{\mathrm{e}}$ (left) and in a $R_{\mathrm{e}}$ region within 0.8-1.0 $R_{\mathrm{e}}$ (middle). Colour coding is by galaxy morphological type, as in  Figure~\ref{fig:Ave_Zslope}. \textit{Right}: Local metallicity Z against dynamically derived local surface mass density; contours cover spaxels with densities greater than 0.3 times the maximum density, for the inner region within 0.2 $R_{\mathrm{e}}$, (solid lines and filled circles as mean points) and $R_{\mathrm{e}}$ region within 0.8 - 1.0 $R_{\mathrm{e}}$, (dotted lines and open circles as mean points). Here different colors correspond to galaxies of different average central metallicity Z; low ($<\mathrm{Z/Z}_{\sun}>_{\mathrm{Inner}}$ $\sim$ 0-0.9; turquoise), intermediate ($<\mathrm{Z/Z}_{\sun}>_{\mathrm{Inner}}$ $\sim$ 0.9-1.8; goldenrod) and high ($<\mathrm{Z/Z}_{\sun}>_{\mathrm{Inner}}$ $\sim$ 1.8-4; purple).}
\label{fig:All_Z_ZA}
\end{center}
\end{figure*}

The right panel of Figure~\ref{fig:Ave_Zslope} shows a clear separation between different galaxy types; as earlier type galaxies (red) are more metal-rich than late type galaxies (blue) and have stronger gradients. This suggests a correlation between a galaxy's average metallicity $<$Z/Z$_{\sun}>$, and linear metallicity gradient $\nabla$Z. To further explore this trend, we can ask what regions of the galaxy are changing to produce the morphological type  -  $\nabla$Z  correlation. 
For each galaxy, we calculate their average Z within two different regions: an $R_{\mathrm{e}}$ region, between 0.8-1.0 $R_{\mathrm{e}}$ (relatively outer); and an inner region, within 0.2 $R_{\mathrm{e}}$. The left panel of Figure~\ref{fig:All_Z_ZA} shows the Z gradient against these two average metallicities.

Comparing the two regions, we see that the average metallicity of the \textit{inner region} (left) shows very similar correlations with $\nabla Z$ as those seen in Figure~\ref{fig:SP_G}. This suggests the central Z is tightly linked to the strength of Z gradients and a galaxy's morphological type. Conversely, the Z gradient shows no correlation with the metallicity of a galaxy's region around the effective radius.
This result is consistent with the average Z \textit{profiles} for galaxies of a given morphological type shown in the right panel of Figure~\ref{fig:Ave_Zslope} - the inner regions are enriched more efficiently with respect to the outskirts in early type galaxies.

\begin{figure*}
\begin{minipage}[t]{0.33\linewidth}
\centering
\includegraphics[width=0.25\textheight, height=0.32\textheight]{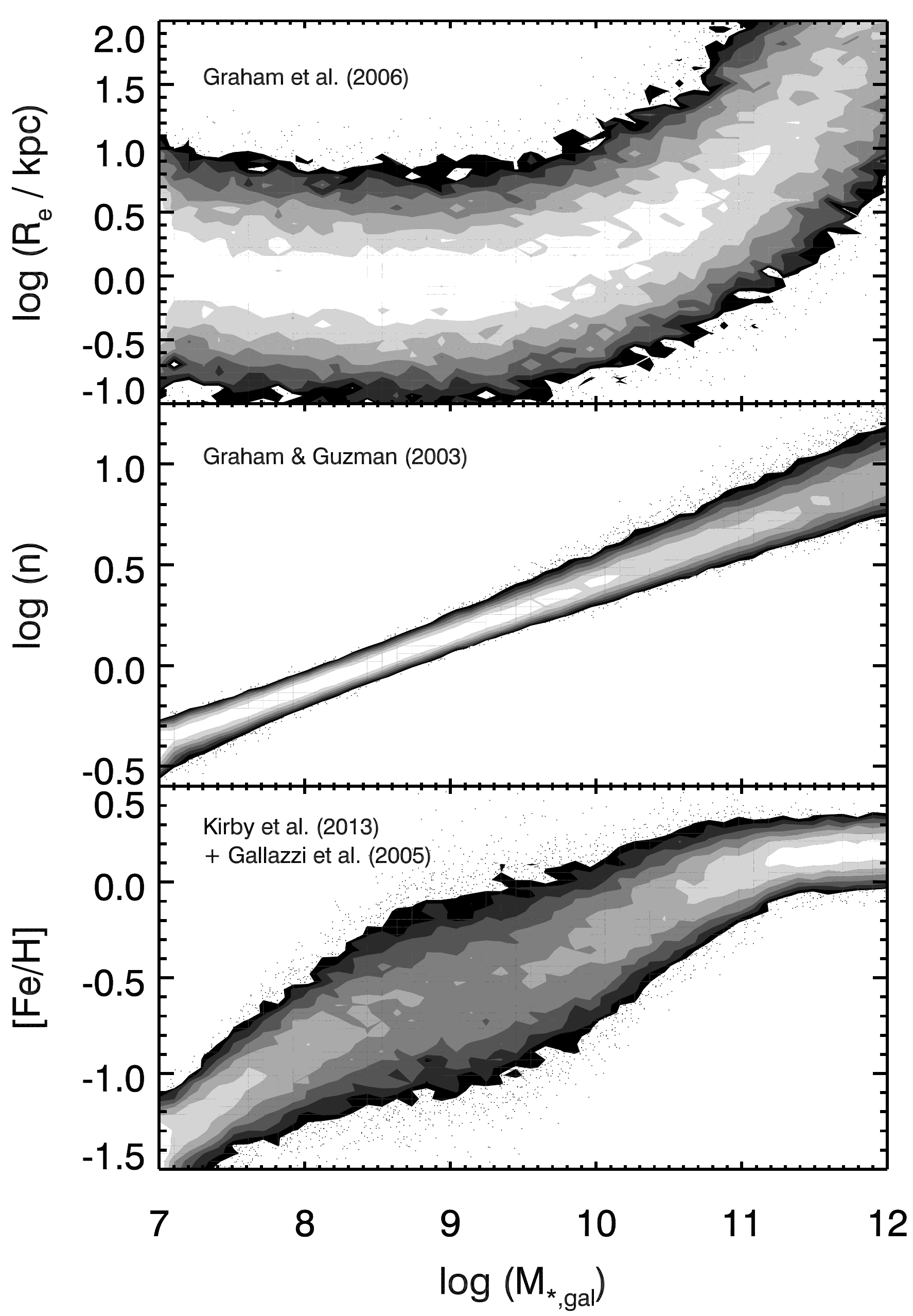}
\end{minipage}%
\begin{minipage}[t]{0.33\linewidth}
\centering
\includegraphics[width=0.23\textheight, height=0.34\textheight]{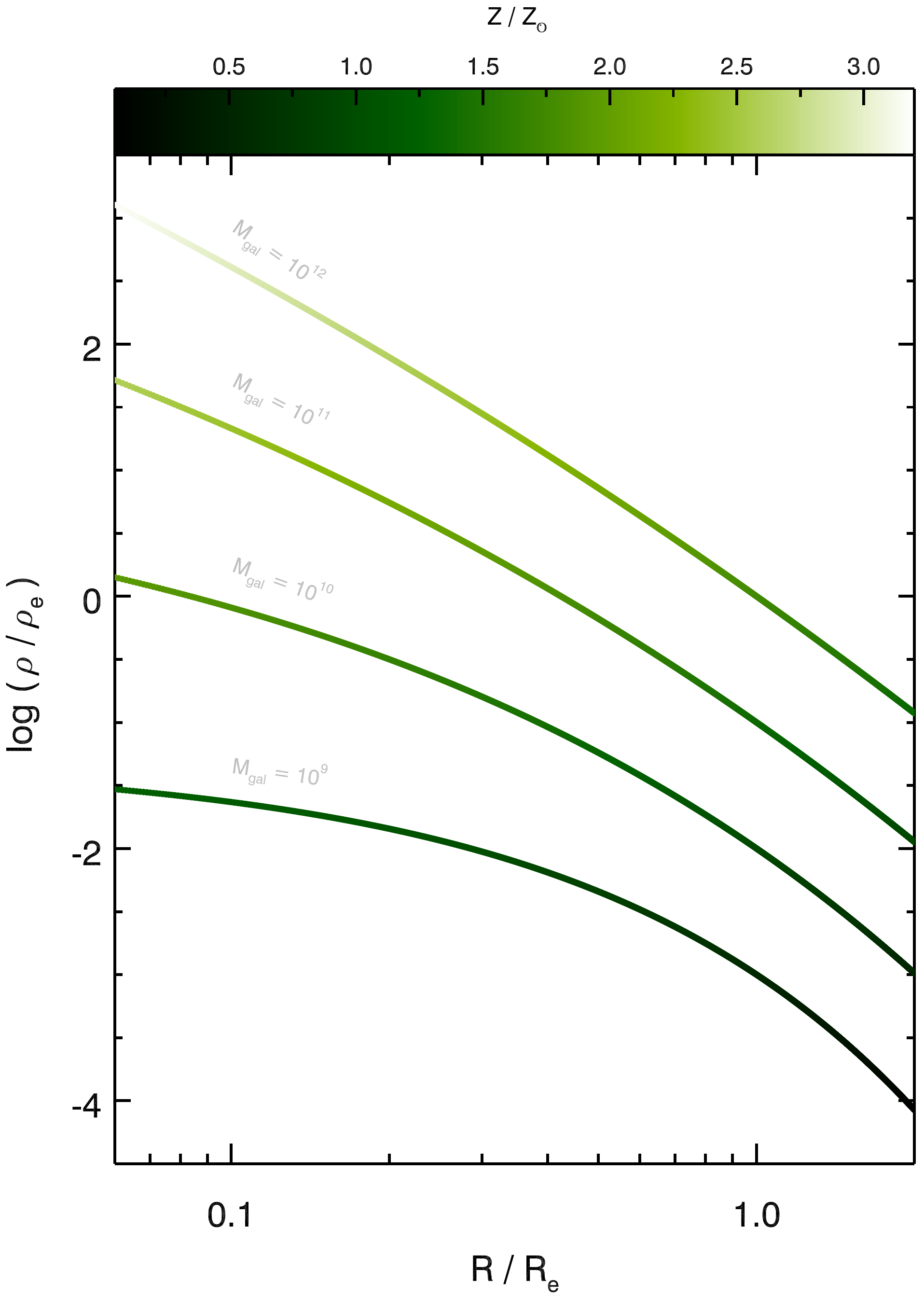}
\end{minipage}
\begin{minipage}[t]{0.33\linewidth}
\centering
\includegraphics[width=0.26\textheight, height=0.34\textheight]{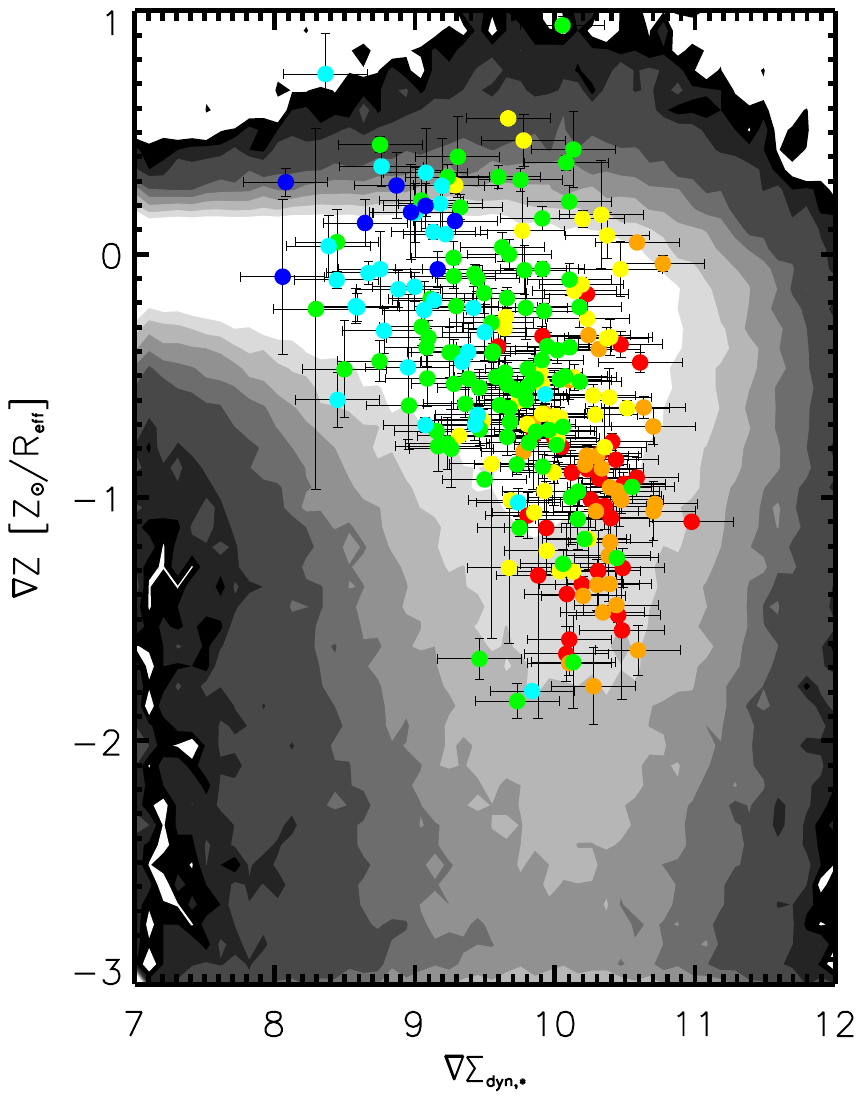}
\end{minipage}
\caption{Left: empirical galaxy scaling relations of size, Sersic index and stellar metallicity, versus galaxy stellar mass.  Middle:  Scaling relations are combined to give representative mass density and metallicity profiles for galaxies of different total stellar masses (color-coded by locally predicted galaxy metallicity).  Right:  Predicted metallicity gradients from the empirical model (contours) naturally accounts for the trend and scatter between  stellar mass density gradient and metallicity gradient seen in our data (points color-coded by galaxy morphological type, as in  Figure~\ref{fig:Ave_Zslope}).}
\label{fig:ZMZ}
\end{figure*}

The interplay between the mass dependence of a galaxy's morphological type, central concentration, and average metallicity allows for an intuitive description of metallicity gradients when viewed in linear Z space.
Previous studies have demonstrated the existence of a \textit{local} mass, gas metallicity, star formation relation~\citep[e.g.,][]{2012ApJ...756L..31R}, hinting that local metal enrichment is the result of integrated star formation/stellar mass accumulation. 

In order to explore a correlation between the local mass density and Z, we divide our sample of galaxies into 3 subgroups according to the average metallicity of their inner regions: (Low: Z $\sim$ 0-0.9, Moderate: Z $\sim$ 0.9-1.8, High: Z $\sim$ 1.8-4). 
For each group, we compare the galaxies' local metallicities in the inner, and $R_{\mathrm{e}}$ regions, with the corresponding dynamically derived stellar surface mass density from Schwarzschild modeling. 
For each subgroup we plot all the spaxels which lie within inner (0.0 $-$ 0.2$R_{\mathrm{e}}$; solid lines) and outer $R_{\mathrm{e}}$ (0.8 $-$ 1$R_{\mathrm{e}}$; dotted lines) regions of all galaxies, in this stellar surface mass density-metallicity parameter space.  The right panel of Figure~\ref{fig:All_Z_ZA} shows density contours (greater than 0.3 times the maximum density)  for spaxels in the outer regions (dotted contours) and inner regions (solid contours) of galaxies with different central metallicities.

These distributions follow a clear \textit{local} mass-metallicity relation, with $R_{\mathrm{e}}$ regions being more metal-poor and having lower surface mass density than the inner regions. Importantly, the Z and surface density \textit{difference} between a galaxy's inner region and $R_{\mathrm{e}}$ regions, increases as the average inner Z (and therefore galaxy mass) increases. 

In this exercise we have used the dynamically derived stellar surface mass density $\Sigma_{dyn,\ast}$, which is computed independent of the stellar populations profiles. Similar results for the correlations between local metallicity and stellar surface mass density are found by replacing $\Sigma_{dyn,*}$ with the stellar surface mass density derived via stellar population synthesis modelling from the available spectro-photometry (Zibetti et al., in prep.).

The right panel of Figure~\ref{fig:All_Z_ZA} implies that as a galaxy's local regions also follow the mass$-$metallicity relation, we would expect a steeper Z gradient when there is a steeper surface mass density gradient. Here we define a dynamically derived stellar surface mass density gradient $\Sigma_{dyn,\ast}$ as:
\begin{equation}
\nabla \Sigma_{dyn,\ast} \equiv {\mathrm{log}}(\Sigma_{dyn,\ast (r=0)}-\Sigma_{dyn,\ast (r=Re)})
\end{equation}
The right panel of Figure~\ref{fig:ZMZ} explicitly quantifies the link between $\nabla \Sigma_{dyn,\ast}$ and stellar Z gradients, and shows that the Z gradients become steeper as the mass gradients increase in magnitude, though with considerable scatter. In general, early type galaxies with steeper Z gradients also have proportionally higher central Z and steeper mass density gradients, reaffirming the relations, between Z gradient, mass concentration and morphological type.

\subsection{Empirical model}

This simple picture, that stellar metallicity gradients are most clearly observationally connected to, and result from, the time-integrated differences in stellar mass buildup between the inner and outer parts of a galaxy, can be crudely tested using independent empirical scaling relations for galaxy populations. We begin by considering observed relations between galaxy stellar mass and light profile~\citep[Sersic index $n$,][]{2003AJ....126.1787G}, stellar metallicity 
\citep{2013ApJ...779..102K}, and effective radius ~\citep{2006AJ....132.2711G}, as shown in the left panel of Figure~\ref{fig:ZMZ}. 

In the following exercise we stochastically draw properties which a galaxy of mass $M_{*}$ would have, from these scaling relations.  The mass dependent relations are given a scatter representative of the observations: $\sigma_{log R_{e}} = 0.37$ dex, $\sigma_{log n} = 0.05$ dex  (van der Wel et al. 2014), and the $1-\sigma$ spread on [Fe/H] as a function of mass directly from \citep{2005MNRAS.362...41G} and \citep{2013ApJ...779..102K}.

For a galaxy of a given mass, we use the effective radius and Sersic index to compute a representative Einasto profile of the 3D stellar mass distribution ($\rho(r)$) for that galaxy. In order to link the global metallicity relations with the local structural relations, at any point $r$ on the radial mass density profile, $\rho(r)$, we compute the mass an idealized spherical galaxy would have if it was this \textit{average} density and size, i.e. $M_{*}(r) \propto \rho(r) r^{3}$. That idealized galaxy mass, and thus the corresponding region at $r$ in the composite model galaxy, is then given a homogeneous metallicity following the observed galaxy global mass-metallicity relation \citep[e.g.,][]{2005MNRAS.362...41G}. The composite galaxy profile is constructed by repeating this exercise at all radii - and the resulting profile can have a metallicity and density gradient computed as per our observations.

The construction of a mock galaxy whose local properties are set by global scaling relations, is physically justified by observations which suggest that the outer regions of disk galaxies are forming stars and enriching chemically in a comparable way to lower mass dwarf galaxies.  Hence the local sub-galaxy which is constructed to have radius $r$, will only represent a local shell of that corresponding metallicity at $r+dr$ in the final composite galaxy.

This simple exercise combines galaxy scaling relations to allow for a coarse estimate of local metallicity, but should not be over-interpreted, as it is essentially relying on dimensional analysis to produce a stochastic prediction for where \textit{populations} of galaxies may lie in the $\nabla \Sigma_{dyn,\ast} - \nabla Z$ plane.

The middle panel of Figure~\ref{fig:ZMZ} shows these representative mass density profiles for galaxies of different masses, and colour coded by local metallicity. The metallicity changes in these profiles are in qualitative agreement with the profiles in Figure \ref{fig:Ave_Zslope}.  The predicted quantitative correlation between mass density gradient and metallicity gradient using only these empirical relations is shown in the right panel of Figure~\ref{fig:ZMZ} as the contours. The scatter in the M$_{*}-n$, M$_{*}-R_{e}$ and mass-metallcity relations accounts for nearly all the observed intrinsic scatter in the $\nabla $Z - $\nabla \Sigma_*$ relation.  We suggest that the larger scatter in the $\nabla$Z  gradient versus total mass (Figure~\ref{fig:FeH_Z_DB}), is a consequence of the fact that the establishment of a metallicity gradient is a \emph{local} phenomena in galaxies. Our empirical model also produces some galaxies with positive slopes, which are observed in high percentage for our late type galaxies.  We will discuss the possible origin of these objects in Section 5.5.

It is rather surprising that such a simple dimensional analysis exercise based on galaxy structural and global chemical relations does a reasonable job in reproducing the average trend and scatter in $\nabla $Z - $\nabla \Sigma_*$.  It is unclear whether this is due to some fundamental simplicity which links local and global galaxy relations, for example a self-similar statistical form for how galaxies chemically enrich (e.g., \citealt{Oey00,Leaman12b,Hartwick15}).  However, it should be kept in mind that the empirical model may be reproducing the broad trends of this parameters space out of coincidence, and the ingredients are not necessarily causal in determining a galaxies structural and chemical gradients.  This must surely be true in some sense for galaxies on an individual scale, given the significant number of mechanisms which are not incorporated into the model (e.g., AGN/star formation driven winds, radial migration, mergers).  Nevertheless, this simple model can provide a starting point for continued and more detailed investigation of such chemodynamical relations.


\subsection{Chemodynamic spatial signatures}
Due to the wealth of kinematic information in CALIFA, and dynamical models for our galaxies (Zhu et al. 2017), we are able to investigate if any dynamical properties appear to be stronger causal factors in setting the stellar population gradients.  
This is motivated by some previous works on low mass galaxies, for example ~\cite{2013MNRAS.434..888S} used simulations to investigate the metallicity gradients of dwarf galaxies. Their results suggested that metallicity gradients can be built up efficiently during the evolution of non-rotating dispersion-dominated dwarf galaxies, as low angular momentum allows more gas to concentrate in a galaxy’s central regions, leading to proportionally more star formation.

While this scenario is in agreement with some local group dwarf galaxies' observations~\citep[e.g.,][]{2013ApJ...767..131L,2015ApJ...798...77H,2017MNRAS.466.2006K}), the gradient differences also span differences in mass SFH and environment as well as angular momentum, making a direct causal link between dynamics and the metallicity gradients unclear. More relevant for this study is the unknown dependence of this mechanism on host galaxy mass.  Given the orbital decomposition that our Schwarzschild models provide, we can directly look for correlated behaviour of chemical and kinematic properties of our sample of higher mass galaxies.

\begin{figure}
\begin{center}
\includegraphics[width=0.5\textwidth]{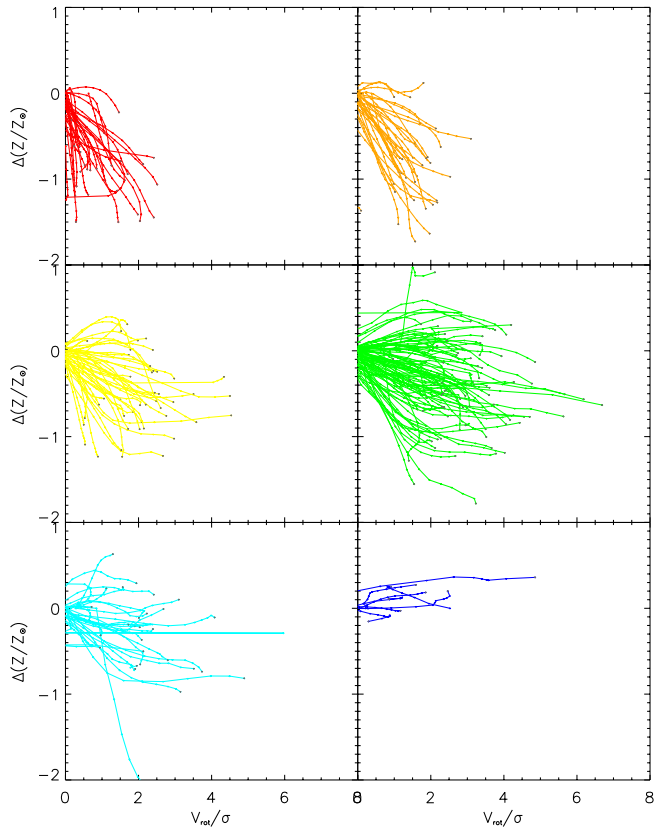}
\caption{Metallicity $\Delta$(Z) against $V_{rot}$/$\sigma$ profiles for different galaxy types, each profile line represents one galaxy's Z and $V_{rot}$/$\sigma$ values at different radii (from 0 to 1 $R_{\rm{e}}$). Color coding is by galaxy morphological type, as in  Figure~\ref{fig:Ave_Zslope}. }
\label{fig:Chemkin_6}
\end{center}
\end{figure}
 
Figure~\ref{fig:Chemkin_6} shows a radial profile for each galaxy's $V_{rot}$/$\sigma$ and $\Delta$Z values, with $V_{rot}$/$\sigma$ increasing from inside to outside.  Different galaxy types show some separation in the figure, as early type galaxies tend to have steeper gradients and low angular momentum, while later type galaxies show an increase in diversity of their tracks, but on average a flatter set of metallicity profiles extending to higher $V_{rot}$/$\sigma$.  Unlike previous studies for dwarf galaxies, the higher mass galaxies likely have a greater number of factors influencing their evolution than the dwarf galaxies (bars, spiral arms, higher merger rates), and indeed a diverse set of tracks is evident.

\begin{figure}
\begin{center}
\includegraphics[width=0.5\textwidth]{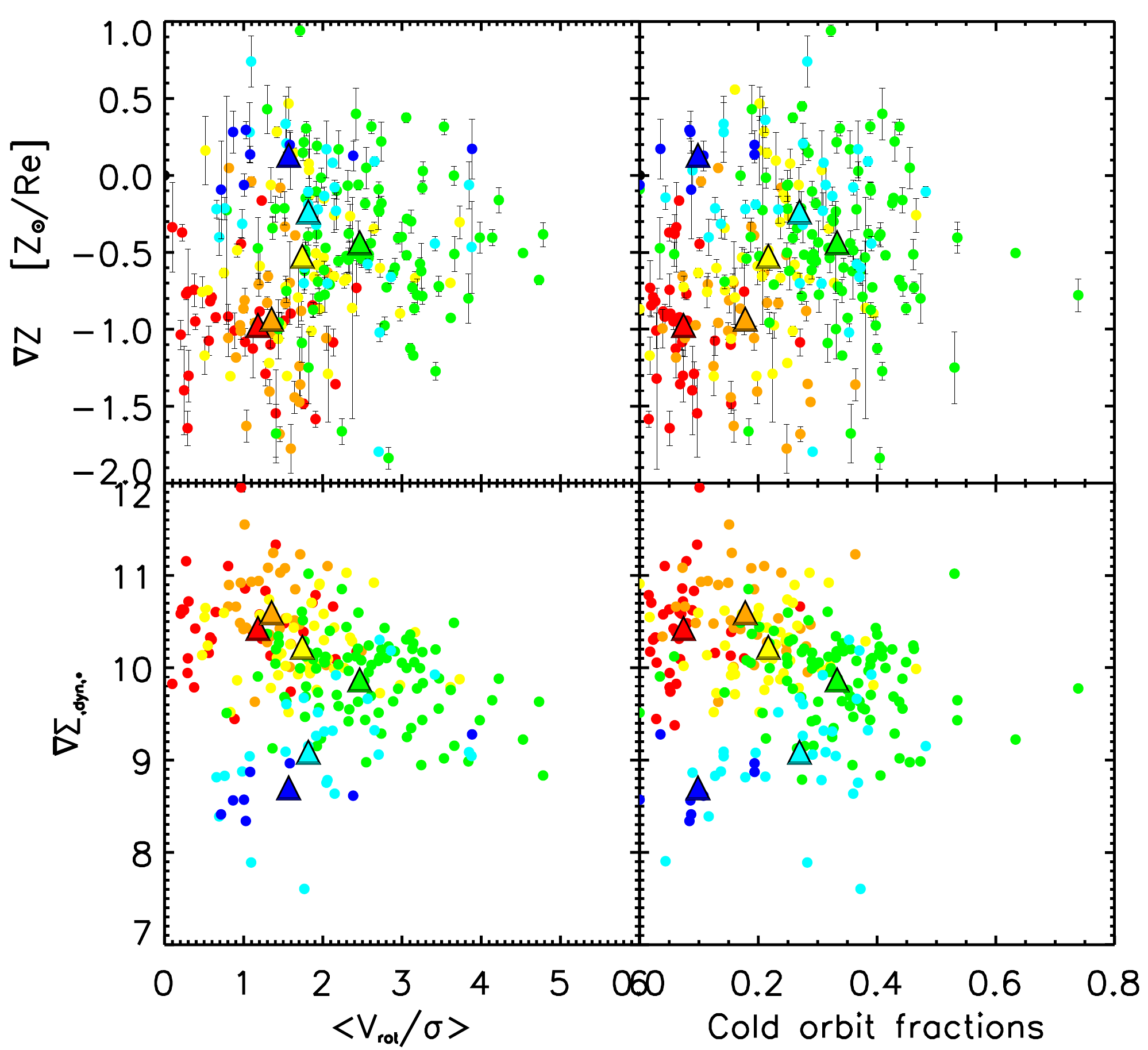}
\caption{Top: Metallicity gradient against galaxies' angular momentum (average $V_{rot}$/$\sigma$ between 0.5-1$R_{\mathrm{e}}$) (left) and cold orbit mass fractions (right) of our 244 GALIFA galaxies. Bottom: Central mass concentration (stellar surface mass density gradient) against galaxies'  angular momentum (left) and cold orbit mass fractions (right), triangles are the mean points of galaxies with each morphological types. Colour coded by galaxy morphological type, as in  Figure~\ref{fig:Ave_Zslope}.}
\label{fig:Vsig_Fraction_Z}
\end{center}
\end{figure}

 \begin{figure*}
\begin{center}
\includegraphics[width=1.0\textwidth, height=0.35\textheight]{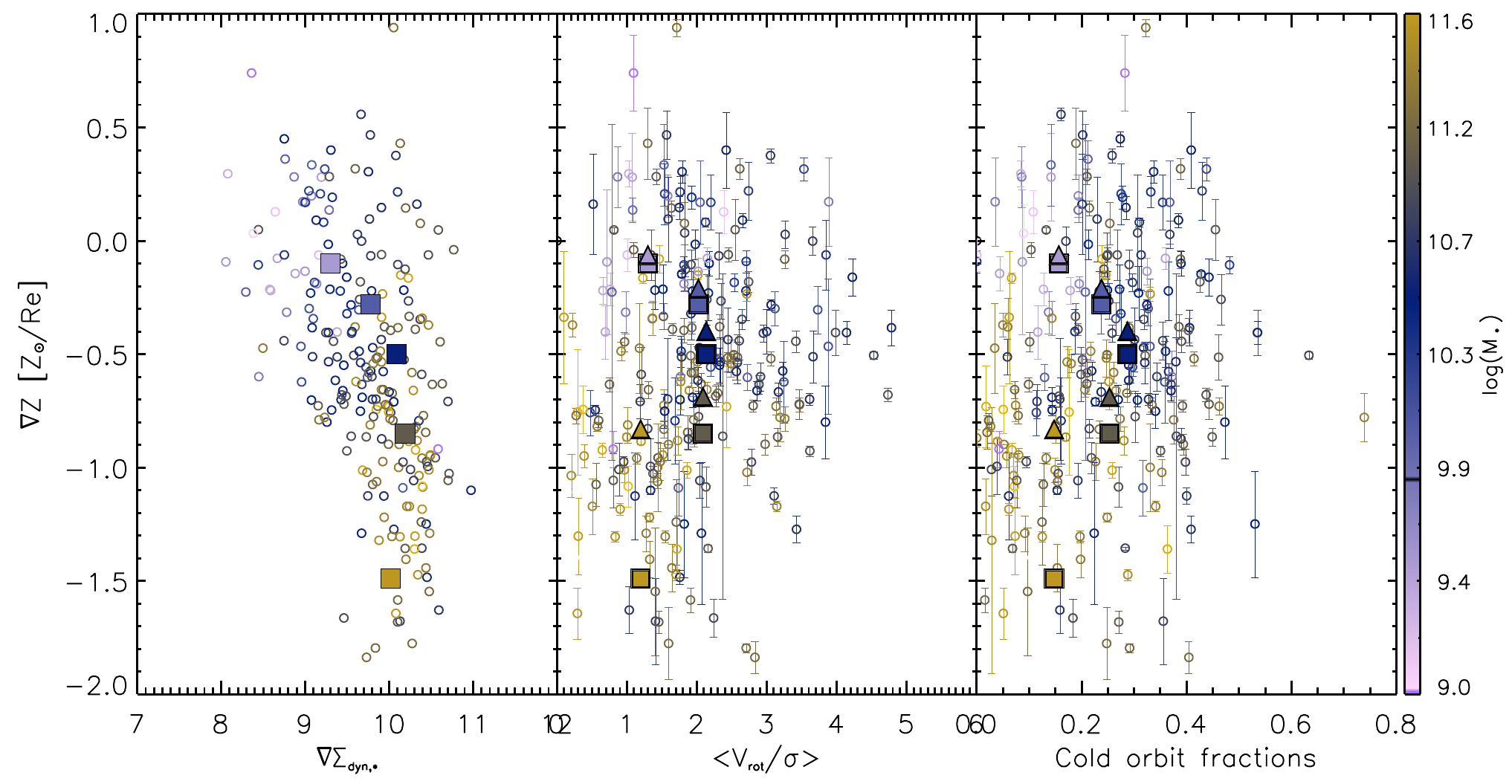}
\caption{A repeat of Figure ~\ref{fig:ZMZ} right panel and the bottom panels of Figure ~\ref{fig:Vsig_Fraction_Z}. Left: shows the correlation between $\nabla \Sigma_{dyn,\ast}$ and $\nabla Z$, Middle and right panels: the matallicity gradient against $V_{rot}$/$\sigma$ and cold orbit fraction, but with color coded by galaxies' stellar mass. Triangles are the mean points of galaxies within each mass bins, blocks are the predictions using the empirical model.}
\label{fig:Vsig_Fraction_Z_M}
\end{center}
\end{figure*}  

To quantify correlations between angular momentum and metallicity gradients in our sample, we use the average $V_{rot}$/$\sigma$ (between 0.5 -1$R_{\mathrm{e}}$) and the disky cold orbit mass fraction as two indicators of a galaxy's angular momentum. The top panels of Figure~\ref{fig:Vsig_Fraction_Z} shows the metallicity gradients against these angular momentum proxies. It is evident that both early and very late type galaxies tend to have low angular momentum, but show different metallicity gradients. 
From type Sd to type Sb galaxies, as the metallicity gradient becomes steeper, the angular momentum increases, until a turning point at Sb. From type Sb to type E,  the metallicity gradient steepens, while the angular momentum starts to decrease. Similar trends are seen in the bottom panels of Figure~\ref{fig:Vsig_Fraction_Z} when replacing metallicity gradients with stellar surface mass density gradients. 

To better understand the role of mass concentration and angular momentum in setting metallicity gradients, we divide our galaxies into 5 groups according to their stellar mass log M$_*$ from 9 to 11.5 with 0.5 dex wide bins. We then get mean locations of each mass bin in $\nabla $Z and $V_{rot}$/$\sigma$ or cold orbit fraction planes (shown as triangles in middle and right panels of Figure~\ref{fig:Vsig_Fraction_Z_M}). With the empirical model we can compute the \emph{average} $\nabla \Sigma_*$ and $\nabla Z$ of model galaxies in the corresponding mass bins, and we show these as squares in the left panel of Figure~\ref{fig:Vsig_Fraction_Z_M}. In general the average mass bins for the model galaxies follow the average observed relation - however for the highest mass bins, the model galaxies show steeper Z gradients (at fixed density gradient) than what are observed in our CALIFA sample.  This could speak to additional factors not included in our simple model (such as AGN driven outflows, or satellite galaxy mergers) altering the density and metallicity profiles of these galaxies.  An interesting test with larger surveys would be to look at an environmental dependence of the $\nabla \Sigma_* - \nabla Z$ high mass regime to see if any systematic movement is apparent

The projections of the average model galaxies into the angular momenta parameter space is shown in the middle and right panels of Figure~\ref{fig:Vsig_Fraction_Z_M}.  As we don't explicitly compute an angular momentum observable for the model galaxies, we have associated the mass bins to the same angular momentum as is observed for galaxies of that mass.
Compared to the observed mean chemo-dynamical values (triangles), the predicted points clearly show a similar mass dependence - suggesting that angular momentum is not a unique predictor of metallicity gradient, and that the primary factor in predicting the gradients should be the mass density profile. 
This is different from previous studies of low mass range (M$_{*}<10^{10.5}$M$_{\sun}$) galaxies, where it was found that higher angular momentum galaxies tended to have steeper gradients. This suggests the role of angular momentum in preventing central gas build-up and the role of outer perturbations may not be a causal formation process for metallicity gradient in higher mass galaxies. The turning point around Sb galaxies in this parameter space is perhaps due to the correlation between angular momentum and mass - as beyond this mass, mergers likely alter the stellar angular momentum in galaxies.  However, we must be careful as the current dynamical state is not necessarily representative of all epochs in the galaxies' evolution.

\begin{figure*}
\begin{center}

\includegraphics[width=0.95\textwidth, height=0.35\textheight]{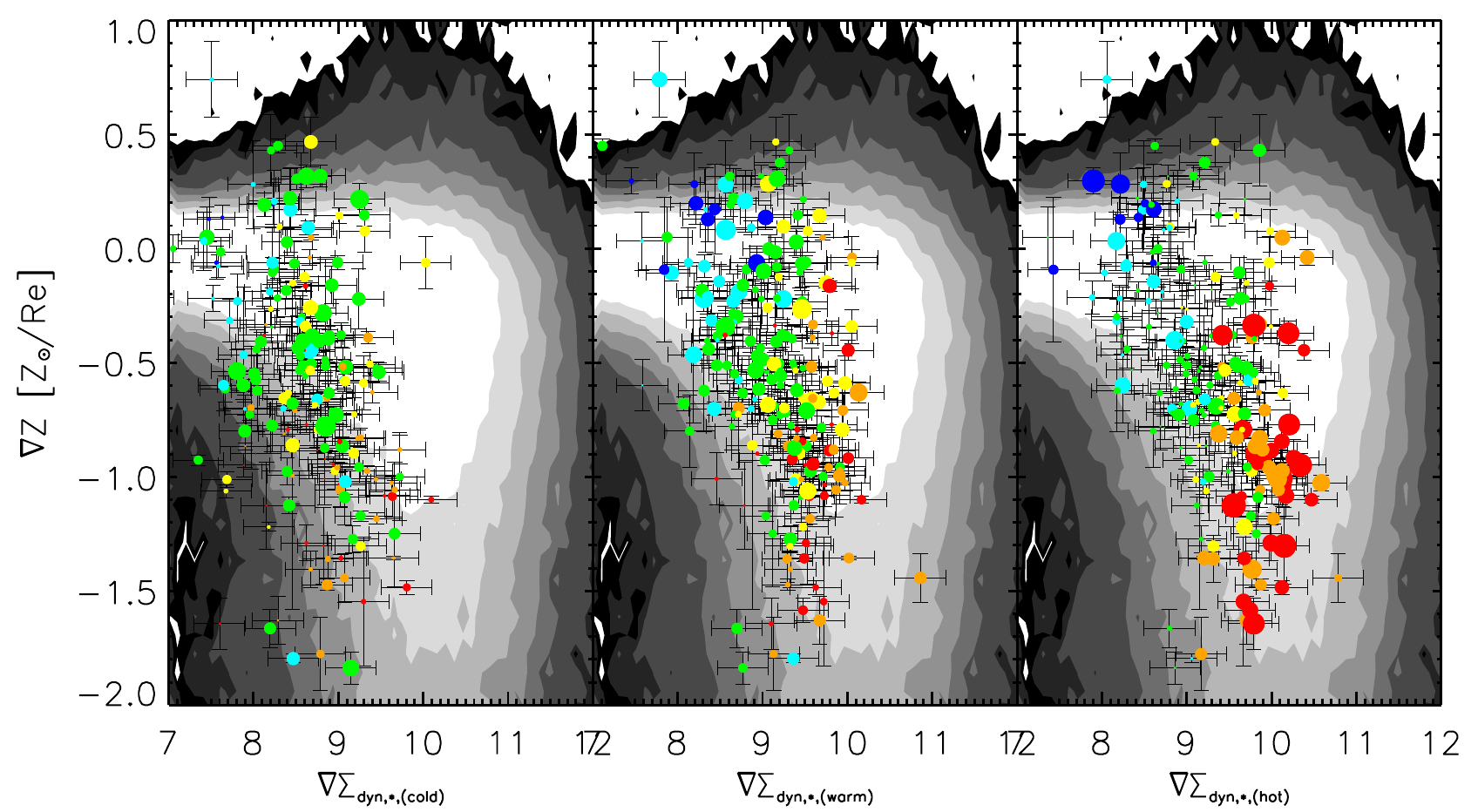}
\caption{Metallicity gradient within 1$R_{\rm{e}}$ against stellar surface mass density gradients of different stellar orbit components: cold orbits $\nabla \Sigma_{dyn,*,(cold)}$ (left), warm orbits $\nabla \Sigma_{dyn,*,(warm)}$ (middle) and hot orbits $\nabla \Sigma_{dyn,*,(hot)}$ (right), . Symbol size represents the mass fraction of the respective components. Grey contours are the total $\nabla $Z - $\nabla \Sigma_*$ relation from our model. Colour coded by galaxy morphological type, as in  Figure~\ref{fig:Ave_Zslope}.}
\label{fig:orbit_Mg_Zslope}
\end{center}
\end{figure*}

From our Schwarzschild models we are in the unique position to study any correlations between metallicity gradient and the surface mass density gradients of different dynamical components - which may give some insight into which epoch or assembly method imparted the present day metallicity profiles. We proceed by computing the stellar surface mass density profiles from \textit{just} the cold, warm or hot orbital components.  Differing behaviors might hint at a time dependence of gradient formation - as hot orbital populations may be composed of the earliest generations of stars, while cold orbital populations represent newly formed young stars. Figure~\ref{fig:orbit_Mg_Zslope} shows the stellar surface mass density gradients of cold  $\nabla \Sigma_{dyn,*,(cold)}$, warm $\nabla \Sigma_{dyn,*,(warm)}$ and hot $\nabla \Sigma_{dyn,*,(hot)}$ orbit components against the metallicity gradients within $R_{\rm{e}}$. 

In Figure~\ref{fig:orbit_Mg_Zslope}, we see a progression where the $\nabla $Z - $\nabla \Sigma_*$ correlation from Figure~\ref{fig:ZMZ}, becomes more evident as we look at hotter orbit densities. It appears that regardless of the final angular momentum of the galaxy, the density profile becomes established in conjunction with the formation or heating of the stars that make-up the hot orbits.  

This could suggest that the density gradients of the coldest component may have large stochastic spatial variations due to young stars forming inhomogeneously out of a self-enriching ISM.  Then over time a superposition of these radial distributions may statistically drive to a more regular $\nabla $Z - $\nabla \Sigma_*$ correlation.  However this again suggests that while there is a clear interconnection between: 1) the stellar mass build-up and dynamical state of a galaxy, and 2) the stellar mass build-up and the metallicity gradient, that it is not yet clear that present day dynamical structure has a causal link with the strength of metallicity gradients - at least within one $R_{\rm{e}}$ and for galaxies in this mass regime.  This link may be observationally difficult to untangle as the hot and warm orbits can simultaneously trace two different processes (e.g., stars born from a dynamically cooling ISM, and stars which were dynamically heated after their birth \citep{Leaman17}.

\subsection{Radial migration}

Several theoretical works have demonstrated that radial migration of stars within a galaxy's disk could be a possible channel to flatten the radial stellar population gradient ~\citep[e.g.,][]{2008ApJ...675L..65R,2016ApJ...818L...6L,2016ApJ...820..131E}. Clear observational signatures of radial migration remain hard to achieve however.  Recently,~\cite{2017arXiv170502120R} studied profiles that extend to the very outer disk regions of a subset of the CALIFA sample presented here, and found that galaxies which have up-bending surface brightness profiles (e.g., disk surface density profiles which increase beyond some radius in the outer regions), show flatter stellar metallicity gradients in these regions - with very small statistical differences. While they focus on the outer regions of disks, given that spiral arms and bars should be found in high mass, cold disk galaxies, we can ask if our sample shows any second-order trends in the metallicity gradients that could be ascribed to radial migration.

\begin{figure*}
\begin{center}
\includegraphics[width=0.9\textwidth]{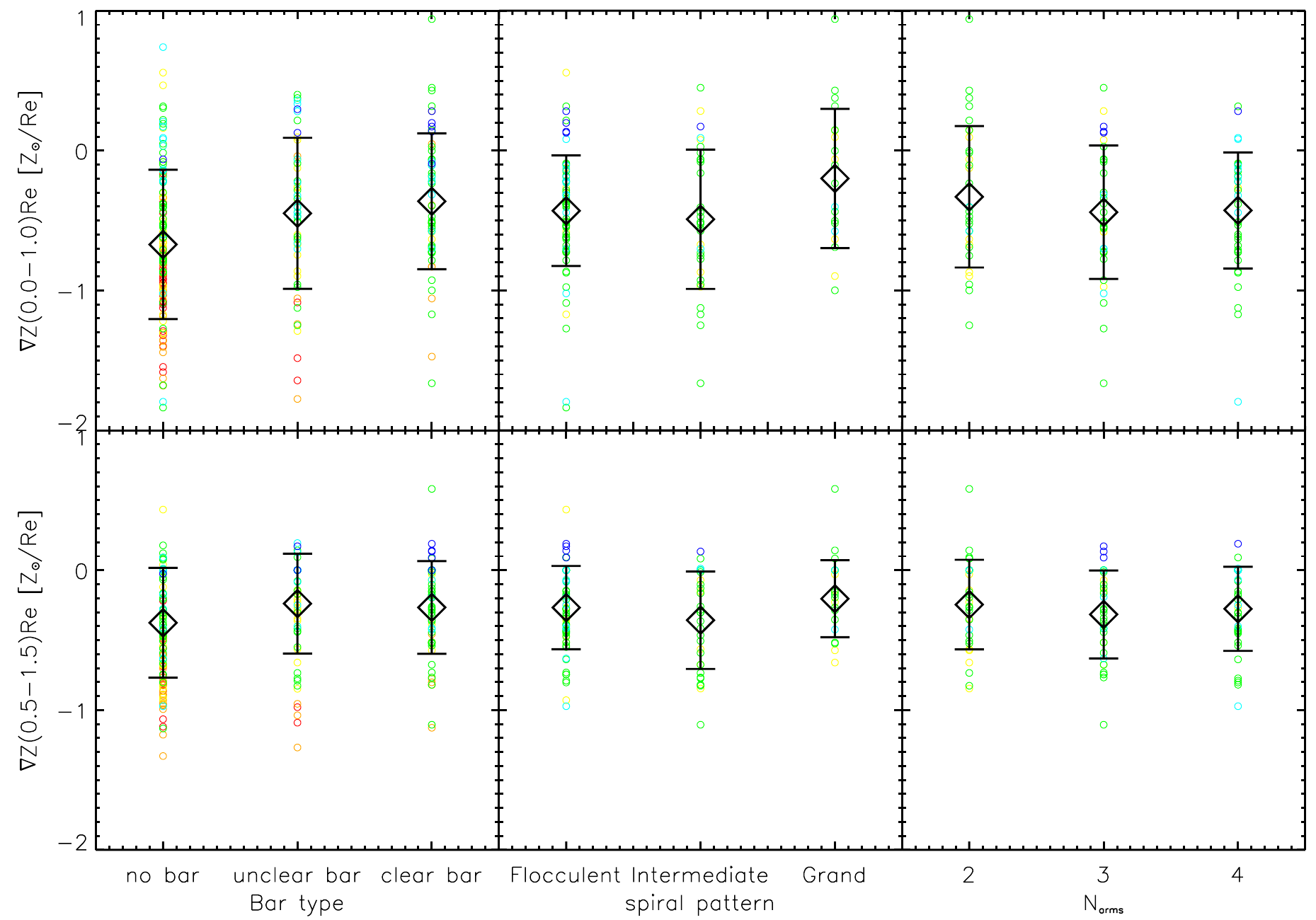}
\caption{Metallicity gradient within 1$R_{\rm{e}}$ (top) and between 0.5 - 1.5$R_{\rm{e}}$ (bottom) against bar type (0: no bar, 1: unclear bar, 2: clear bar)  of our 244 CALIFA galaxies (left),  spiral pattern (1: flocculent, 2: intermediate, 3: grand designed) (middle) and the number of spiral arms (left) of all spiral galaxies. Color-coded by galaxy morphological type, as in  Figure~\ref{fig:Ave_Zslope}.}
\label{fig:Z_Sub}
\end{center}
\end{figure*}

To quantify this, in the top left panel of Figure~\ref{fig:Z_Sub} we show our stellar metallicity gradients against bar strength classifications of galaxies in our sample. There is a correlation between bar type and metallicity gradient in that galaxies with a clear bar feature show flatter Z gradients on average, but the correlation is quite weak compared to the scatter  (see also Ruiz-Lara et al., 2017). 
The top middle and right panels explore the relationship between the metallicity gradient and the type of spiral pattern, and the number of spiral arms for the spiral galaxies in our sample. A signature might be expected here as grand design or two-armed spirals should have proportionally more mass in the arms, leading to a stronger scattering of stars in the disc. Just like the bar-stellar metallicity gradient relation in the left panel, we also see weak correlations between the spiral pattern and the diversity of metallicity gradients (grand design galaxies are averagely flatter in Z gradients).  

As the radial migrations may have different impacts on the inner and outer regions of the disk, in the bottom three panels we also show the stellar metallicity gradient between 0.5-1.5Re against bar strength, spiral pattern and spiral arm number. Excluding the very central region the trend does not change, but the correlations appear even weaker.

We note that we also do see a weak correlation between metallicity gradient and any of the higher order terms in our kinematic harmonic decomposition (which would have indicated radial non-circular motions).


Our sample shows that the most apparent observational metrics of radial migration (bars, thin disk orbits, spiral arms) do not strongly correlate with the stellar population gradients within 1$R_{\rm{e}}$. This indicates that radial migration might be a secondary effect in the inner regions, but the best predictor of a galaxy gradient is still the present day mass density profile.

\subsection{Positive metallicity gradients}

\begin{figure}
\begin{center}
\includegraphics[width=0.5\textwidth]{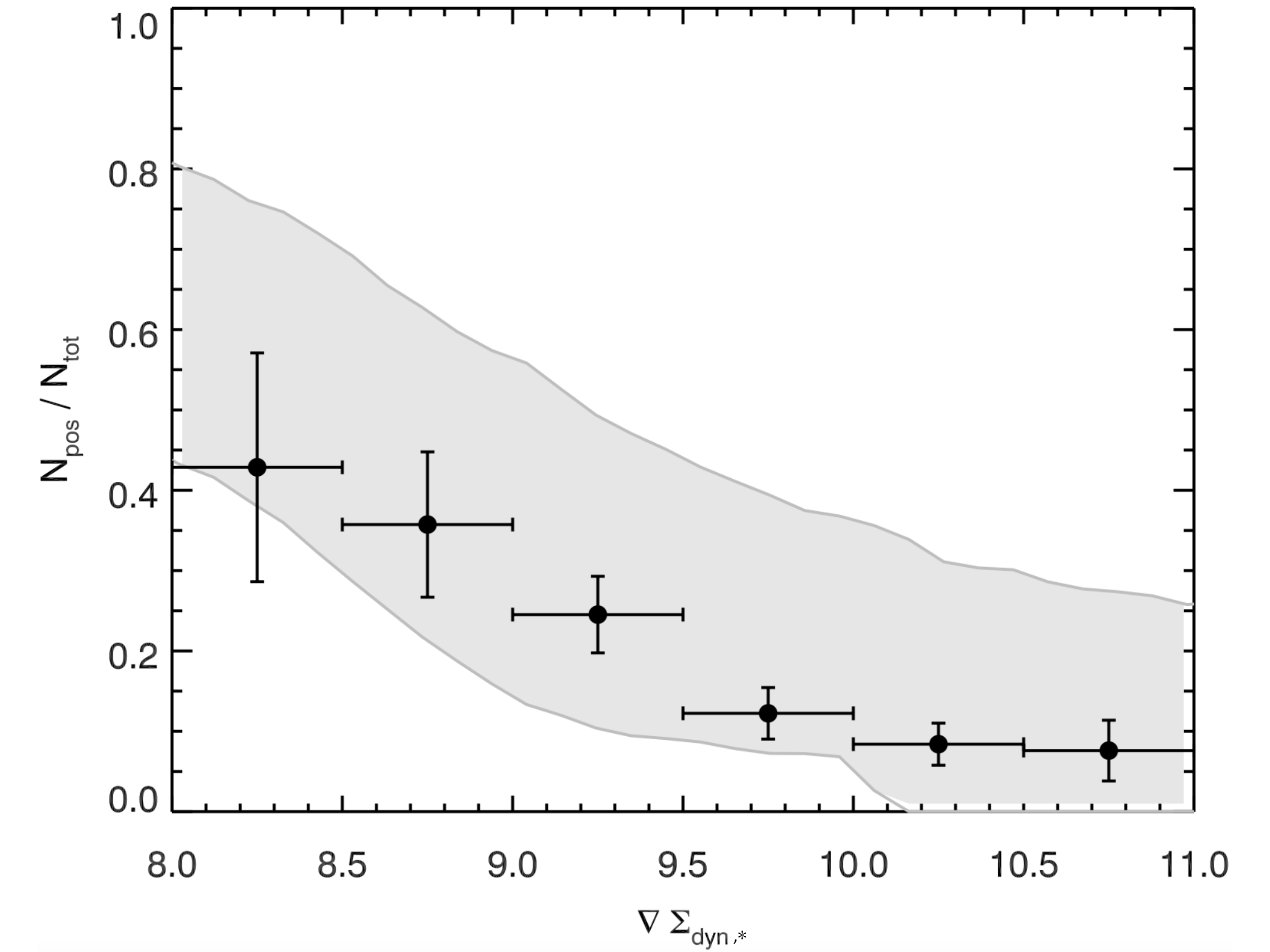}
\caption{Fraction of galaxies with positive metallicity gradients as a function of the dynamically derived stellar mass density gradient.  Grey band shows the predictions from the simple empirical model in Figure \ref{fig:ZMZ}, black points show the fraction of positive metallicity gradients amoung the CALIFA galaxies.}
\label{fig:Z_pos}
\end{center}
\end{figure}

Our sample shows a number of galaxies with positive metallicity gradients, with 12\% of all of our galaxies having metallicity gradients with $\nabla Z >$0.1. We find that late type galaxies are more likely to show positive metallicity gradients, as from type E to Sd the fraction of positive gradients rises from: 0\%, 0\%, 11.1\%, 13.8\%,  19.3\% and 75.0\%. 
In the context of our empirical model, these galaxies result from the scatter in metallicity at fixed local mass being large enough to allow for the formation of positive metallicity slopes.  We show in Figure \ref{fig:Z_pos} the predicted fraction of galaxies in the simple model as a function of galaxy density gradient.  The simple model of Figure \ref{fig:ZMZ} predicts the fraction of positive gradients rises as galaxies exhibit flatter density profiles, in excellent agreement with our data.

The underlying physics for the scatter in metallicity, or the shallowness of some density profiles is likely more complex than the simple model however. Some previous studies have explored the possible scenarios which may be relevant for these positive gradients galaxies. For example, ~\cite{2016ApJ...820..131E} used simulations to explore the stellar feedback processes in low-mass galaxies and suggested that strong feedback could effectively cause a radial migration by strongly perturbing the gravitational potential. 
In addition, by blowing out the central metal-rich gas, the intense feedback may lead to relatively low central metallicities (with respect to their 1$R_{\rm{e}}$ region) compared to other galaxies ~\citep[see also][]{2007AAS...21111102B,2013A&A...554A..47G,2017MNRAS.467.1154S}.  

Stellar feedback is expected to be proportionally more impactful in low mass galaxies. This could mean that more metals are lost due to feedback-driven outflows in low mass galaxies, due to their shallower potential wells.
Concurrently, the star formation efficiency might be reduced in these low mass galaxies due to the impact of feedback on the ISM~\citep{2007ApJ...655L..17B}. Both effects will be enhanced if the intrisically lower angular momentum of lower mass, late type galaxies helps gas funnel to their central regions.
Together, these feedback effects would work to suppress metal build-up in the central regions, shallow the density profiles, and cause larger scatter in local metal enrichment -  with increasing efficiency towards lower mass galaxies. This mechanism would be in qualitative agreement with our finding of more positive gradients as galaxy mass decreases (or moves to later types; Appendix A).  

\section{Conclusions}

In this paper, we study the radial stellar metallicity gradients of 244 CALIFA galaxies, spanning galaxy morphological type from E to Sd. We use a stellar kinematic based ellipse systems to extract flux-weighted stellar population profiles, and use a harmonic decomposition method to derive the kinematic profiles within 1$R_{\rm{e}}$. As our sample covers a broad range of mass/morphological type galaxies, in order to avoid the effect of logarithmic suppression, we study their linear metallicity (metal mass fraction; Z) profiles and gradients.

$\bullet$ We find linear metallicity gradients $\nabla$Z show clear galaxy morphological type dependence, with early-type galaxies tending to have steeper linear metallicity gradients $\nabla $Z. The direct reason behind the Z gradient-G type correlation is the high central metallicity in earlier type galaxies. 

$\bullet$ To further understand the origin of this correlation, we split our galaxies into three different groups according to their central Z, and test the local mass-metallicity relation at different radii in different groups of galaxy. The result not only shows a clear local mass-metallicity relation but also suggests a correlation between surface mass density gradient and $\nabla $Z . 

$\bullet$ We show that with empirical mass-dependent sersic index, size and metallicity scaling relations, the dependence and observed scatter of Z gradients and local stellar mass density can be reproduced.

$\bullet$ We also examine the role of angular momentum by studying $V_{rot}$/$\sigma$ and cold disc fraction. For higher mass galaxies ($>10^{10.5}$M$_{\sun}$) their $V_{rot}$/$\sigma$ and thin disc fraction are correlated with both Z and stellar surface mass density gradients, but for galaxies with stellar mass below $10^{10.5}$M$_{\sun}$ their angular momentum is anti-correlated with the steepness of metallicity/surface mass density profiles.  Analysis of the mass density in different orbital components further suggests angular momentum itself is not directly causal in setting the stellar metallicity gradient of the inner regions of galaxies.

$\bullet$ In order to explore the influence of radial migration, we use observable substructures (bar, spiral arm) as tracers of radial migrations. We find galaxies' radial migration features don't show strong correlations with stellar population gradients, which suggests the radial migration may only provide secondary influence on the shape of gradients, at least within 1$R_{\rm{e}}$.

$\bullet$ Finally, around 12\% of our galaxies show positive metallicity gradients -however the fraction increases to as high as $\sim 80\%$ as galaxy mass decreases. We suggest stellar feedback could be one of the most possible reasons behind this trend.

Though we found the local mass build-up is crucial to a galaxy's metallicity distribution, work still remains to incorporate detailed knowledge about the local star formation history in different regions of a galaxy.  

In this work, our gradients are extracted from metallicity maps, in which the metallicity and mass of a single spaxel is the average metallicity and accumulated mass of all stellar populations it contains. Here we do not take age into our consideration.
There may be interesting correlations between age and metallicity, given that the local mass buildup somewhat reflects the star formation history (e.g., light weighted age).  This formation history may vary in different regions of a galaxy (e.g., bulge vs. disk), and further study will expand on the link between mass buildup and age.
As our results show clear correlations between local mass and metallicity but with some scatter, the local SFH may contribute a possible source of this scatter.

Recently, ~\cite{2017MNRAS.468.1902Z} studied the age distribution of CALIFA galaxies and found a bimodal local-age distribution, with an old and a young peak primarily due to regions in early-type galaxies and star-forming regions of spirals, respectively. In future works, we will test the $\nabla $Z dependence on age using simulations to further explore how different mass build-up processes impart a signature on the present day stellar populations.

\section*{Acknowledgement}  
We thank the anonymous referee for comments which greatly helped the paper. The authours would like to thank Kristian Denis Ehlert, Dainel Rahner, Ignacio Martin-Navarro, Brad Gibson, Sebastian Bustamante and Hector Hiss for insightful comments which greatly helped this manuscript.
This study uses data provided by the Calar Alto Legacy Integral Field Area (CALIFA) survey (http://califa.caha.es/).
Based on observations collected at the Centro Astronómico Hispano Alemán (CAHA) at Calar Alto, operated jointly by the Max-Planck-Institut f\"ur  Astronomie and the Instituto de Astrofísica de Andalucía (CSIC).

RL acknowledges support for this work from funding from the Natural Sciences and Engineering Research Council of Canada PDF award.
GvdV acknowledges support from the Sonderforschungsbereich SFB 881 "The Milky Way System" (subprojects A7 and A8) funded by the German Research Foundation, and funding from the European Research Council under the European Union’s Horizon 2020 research and innovation programme under grant agreement no. 724857 (Consolidator grant ArcheoDyn). 
SZ has been supported by the EU Marie Curie Career Integration Grant "SteMaGE" Nr. PCIG12-GA-2012-326466  (Call Identifier: FP7-PEOPLE-2012 CIG) and by the INAF PRIN-SKA 2017 program 1.05.01.88.04.

\bibliography{revised_version}{}
\bibliographystyle{mn2e}

\appendix
\section{Density-Metallicity Profiles} 
\begin{figure}
\begin{center}
\includegraphics[width=0.5\textwidth]{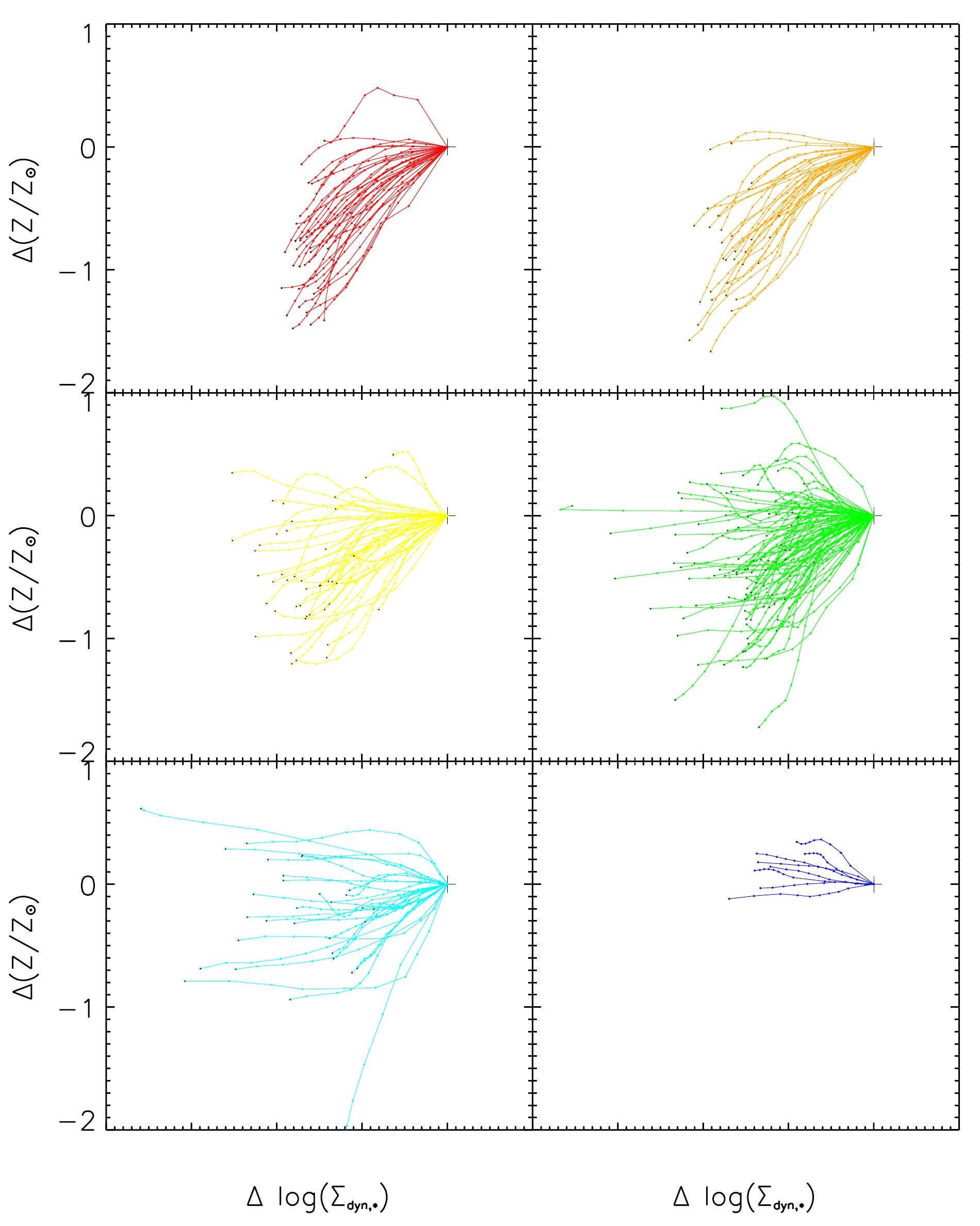}
\caption{Metallicity $\Delta$(Z) against logarithmic dynamically derived stellar surface mass density $log(\Sigma_{dyn,\ast})$ profiles for different galaxy types, each profile represents one galaxy's Z and $\Sigma_{dyn,\ast}$ values at different radii (from 0 to 1 $R_{\rm{e}}$). Color coding is by galaxy morphological type, as in  Figure~\ref{fig:Ave_Zslope}. }
\label{fig:Z_DM}
\end{center}
\end{figure}

Here we show in Figure \ref{fig:Z_DM} the total dynamical mass profiles plotted against the linear metallicity profiles.  Progressing from earlier to later type galaxies we see a larger diversity in their density-metallicity tracks, with late type galaxies showing a larger fraction of positive metallicity profiles.  This is consistent with the results in Figure \ref{fig:Z_pos}, as these late type galaxies tend to have shallower density profiles on average.  However it is clear that there are some objects in Figure~\ref{fig:Z_DM}, with positive gradients but which show significant negative density gradients - which may make these objects interesting test cases for further studies on the impact of feedback on the redistribution of gas and stars.

\begin{table*}
\caption{All galaxies}
\label{table:pars}
\begin{tabular}{@{}lcccccc}

\hline
Galaxy & G Type & log(Mstar/M$_{\sun}$) & $R_{\mathrm{e}}$/kpc & $<V_{rot}$/$\sigma>$ & Z gradient +- err & log(Z) gradient +- err \\
\hline
\hline

          IC0480 &   Sc & 10.148 &  7.631 &  3.878 &  -0.466 +- 0.298 &  -0.227 +- 0.145 \\
         IC0540 &   Sa &  9.875 &  2.045 &  2.722 &  -0.602 +- 0.058 &  -0.305 +- 0.026 \\
         IC0674 &   Sa & 10.908 &  5.153 &  1.634 &   0.145 +- 0.034 &   0.042 +- 0.009 \\
         IC0944 &   Sa & 11.260 &  9.219 &  2.014 &  -0.151 +- 0.020 &  -0.036 +- 0.005 \\
         IC1151 &   Sc &  9.847 &  3.341 &  1.815 &  -0.189 +- 0.032 &  -0.174 +- 0.034 \\
         IC1199 &   Sb & 10.669 &  6.885 &  4.528 &  -0.505 +- 0.017 &  -0.179 +- 0.006 \\
         IC1256 &   Sb & 10.301 &  5.885 &  3.062 &  -0.282 +- 0.024 &  -0.101 +- 0.008 \\
         IC1528 &   Sb & 10.144 &  6.305 &  2.705 &  -0.091 +- 0.049 &  -0.042 +- 0.027 \\
         IC1652 &   S0 & 10.612 &  3.978 &  1.582 &  -0.698 +- 0.268 &  -0.184 +- 0.065 \\
         IC1683 &   Sb & 10.589 &  4.392 &  3.052 &   0.376 +- 0.031 &   0.164 +- 0.013 \\
         IC1755 &   Sb & 10.926 &  6.298 &  1.928 &  -0.383 +- 0.027 &  -0.105 +- 0.008 \\
         IC2101 &   Sc & 10.240 &  8.007 &  2.165 &  -0.230 +- 0.119 &  -0.138 +- 0.079 \\
         IC2247 &   Sa & 10.511 &  6.403 &  1.567 &   0.467 +- 0.107 &   0.170 +- 0.041 \\
         IC4566 &   Sb & 10.954 &  6.123 &  2.526 &  -0.502 +- 0.014 &  -0.167 +- 0.005 \\
         IC5309 &   Sc & 10.276 &  4.952 &  2.058 &  -0.702 +- 0.029 &  -0.428 +- 0.021 \\
         IC5376 &   Sb & 10.655 &  5.799 &  1.408 &  -1.677 +- 0.190 &  -0.518 +- 0.045 \\
  MCG-01-54-016 &   Sc &  9.026 &  4.926 &  1.094 &   0.740 +- 0.167 &   1.022 +- 0.078 \\
  MCG-02-02-030 &   Sb & 10.369 &  4.739 &  2.014 &  -0.615 +- 0.057 &  -0.311 +- 0.028 \\
  MCG-02-51-004 &   Sb & 10.680 &  6.967 &  3.154 &  -0.865 +- 0.040 &  -0.262 +- 0.012 \\
        NGC0001 &   Sb & 10.800 &  3.985 &  2.298 &  -0.524 +- 0.027 &  -0.221 +- 0.012 \\
        NGC0023 &   Sb & 11.040 &  5.509 &  1.438 &  -0.999 +- 0.193 &  -0.451 +- 0.099 \\
        NGC0036 &   Sb & 10.896 &  9.189 &  3.057 &  -0.519 +- 0.030 &  -0.165 +- 0.010 \\
        NGC0155 &    E & 11.176 &  6.796 &  1.594 &  -0.898 +- 0.000 &  -0.271 +- 0.000 \\
        NGC0160 &   Sa & 11.030 &  8.068 &  1.476 &  -0.974 +- 0.082 &  -0.382 +- 0.020 \\
        NGC0171 &   Sb & 10.721 &  7.217 &  1.781 &  -0.512 +- 0.080 &  -0.216 +- 0.032 \\
        NGC0180 &   Sb & 10.922 & 10.450 &  3.615 &  -0.926 +- 0.023 &  -0.315 +- 0.008 \\
        NGC0192 &   Sa & 10.848 &  6.465 &  2.299 &  -0.672 +- 0.039 &  -0.198 +- 0.012 \\
        NGC0214 &   Sb & 10.823 &  5.706 &  3.104 &  -1.125 +- 0.035 &  -0.426 +- 0.011 \\
        NGC0216 &   Sd &  9.282 &  2.225 &  1.004 &  -0.062 +- 0.073 &  -0.078 +- 0.082 \\
        NGC0217 &   Sa & 11.097 &  6.379 &  2.143 &  -0.656 +- 0.033 &  -0.258 +- 0.013 \\
        NGC0234 &   Sc & 10.653 &  6.235 &  3.415 &  -0.443 +- 0.047 &  -0.261 +- 0.038 \\
        NGC0237 &   Sc & 10.310 &  4.452 &  1.759 &  -0.701 +- 0.065 &  -0.336 +- 0.031 \\
        NGC0257 &   Sc & 10.794 &  7.932 &  2.705 &  -1.796 +- 0.019 &  -1.297 +- 0.007 \\
        NGC0364 &    E & 10.962 &  5.617 &  2.129 &  -1.085 +- 0.089 &  -0.276 +- 0.019 \\
        NGC0429 &   Sa & 10.794 &  2.472 &  1.204 &  -0.589 +- 0.119 &  -0.140 +- 0.026 \\
        NGC0444 &   Sc &  9.867 &  7.830 &  2.045 &   0.170 +- 0.184 &   0.144 +- 0.142 \\
        NGC0477 &   Sb & 10.497 &  8.846 &  3.660 &  -0.511 +- 0.114 &  -0.225 +- 0.052 \\
        NGC0499 &    E & 11.401 &  6.468 &  0.388 &  -0.949 +- 0.050 &  -0.230 +- 0.013 \\
        NGC0504 &   S0 & 10.470 &  2.550 &  1.336 &  -0.995 +- 0.124 &  -0.215 +- 0.027 \\
        NGC0517 &   S0 & 10.822 &  2.994 &  1.033 &  -1.628 +- 0.104 &  -0.361 +- 0.020 \\
        NGC0528 &   S0 & 10.874 &  4.100 &  2.059 &  -0.630 +- 0.046 &  -0.154 +- 0.011 \\
        NGC0529 &    E & 11.088 &  4.226 &  0.966 &  -0.445 +- 0.042 &  -0.099 +- 0.010 \\
        NGC0551 &   Sb & 10.641 &  7.134 &  2.606 &  -0.441 +- 0.084 &  -0.151 +- 0.028 \\
        NGC0681 &   Sa & 10.492 &  3.717 &  0.546 &  -0.746 +- 0.022 &  -0.331 +- 0.012 \\
        NGC0755 &   Sc &  9.380 &  3.283 &  0.656 &  -0.219 +- 0.058 &  -0.453 +- 0.089 \\
        NGC0768 &   Sc & 10.542 &  7.483 &  2.564 &  -0.576 +- 0.036 &  -0.204 +- 0.012 \\
        NGC0774 &   S0 & 10.924 &  3.918 &  1.522 &  -0.333 +- 0.058 &  -0.097 +- 0.017 \\
        NGC0776 &   Sb & 10.694 &  6.473 &  2.240 &  -1.663 +- 0.084 &  -0.455 +- 0.040 \\
        NGC0810 &    E & 11.552 &  9.281 &  2.424 &  -0.731 +- 0.179 &  -0.207 +- 0.050 \\
        NGC0825 &   Sa & 10.422 &  2.886 &  0.517 &   0.162 +- 0.222 &   0.047 +- 0.068 \\
\hline
\end{tabular}

\end{table*}

\begin{table*}
\contcaption{All positive galaxies}
\label{table:pars}
\begin{tabular}{@{}lcccccc}

\hline
Galaxy & G Type & log(Mstar/M$_{\sun}$) & $R_{\mathrm{e}}$/kpc & $<V_{rot}$/$\sigma>$ & Z gradient +- err & log(Z) gradient +- err \\
\hline
\hline

        NGC0932 &   S0 & 10.964 &  5.273 &  1.325 &  -1.405 +- 0.079 &  -0.451 +- 0.028 \\
        NGC1056 &   Sa & 10.020 &  1.529 &  0.475 &  -0.757 +- 0.100 &  -0.380 +- 0.041 \\
        NGC1167 &   S0 & 11.692 &  8.520 &  1.462 &  -0.080 +- 0.060 &  -0.026 +- 0.016 \\
        NGC1349 &    E & 10.928 &  7.777 &  1.908 &  -1.584 +- 0.053 &  -0.480 +- 0.016 \\
        NGC1542 &   Sa & 10.438 &  4.096 &  2.064 &  -1.290 +- 0.312 &  -0.482 +- 0.093 \\
        NGC1645 &   S0 & 10.831 &  4.409 &  1.651 &  -0.390 +- 0.027 &  -0.108 +- 0.007 \\
        NGC1677 &   Sc &  9.581 &  2.429 &  1.076 &   0.281 +- 0.193 &   0.330 +- 0.248 \\
        NGC2347 &   Sb & 10.940 &  5.741 &  2.780 &  -0.976 +- 0.058 &  -0.376 +- 0.021 \\
        NGC2449 &   Sa & 10.862 &  5.641 &  2.975 &  -0.897 +- 0.047 &  -0.263 +- 0.015 \\
        NGC2481 &   S0 & 10.684 &  1.440 &  1.190 &  -0.708 +- 0.075 &  -0.168 +- 0.017 \\
        NGC2487 &   Sb & 10.771 &  9.417 &  2.804 &  -0.727 +- 0.025 &  -0.273 +- 0.009 \\
        NGC2513 &    E & 11.539 &  8.628 &  0.204 &  -1.035 +- 0.096 &  -0.262 +- 0.022 \\
        NGC2540 &   Sb & 10.521 &  6.200 &  3.263 &   0.028 +- 0.060 &   0.020 +- 0.032 \\
        NGC2553 &   Sb & 10.838 &  3.079 &  1.445 &  -0.957 +- 0.070 &  -0.227 +- 0.011 \\
        NGC2554 &   S0 & 11.213 &  5.552 &  0.997 &  -0.866 +- 0.016 &  -0.252 +- 0.006 \\
        NGC2592 &    E & 10.618 &  1.362 &  1.337 &  -1.100 +- 0.016 &  -0.284 +- 0.002 \\
        NGC2604 &   Sd &  9.659 &  3.825 &  0.866 &   0.282 +- 0.135 &   0.366 +- 0.179 \\
        NGC2639 &   Sa & 11.168 &  3.927 &  1.891 &  -0.660 +- 0.013 &  -0.164 +- 0.003 \\
        NGC2730 &   Sc & 10.117 &  6.536 &  1.815 &  -0.106 +- 0.037 &  -0.087 +- 0.037 \\
        NGC2880 &    E & 10.671 &  1.940 &  1.755 &  -1.484 +- 0.031 &  -0.424 +- 0.008 \\
        NGC2906 &   Sb & 10.391 &  2.876 &  3.424 &  -1.272 +- 0.058 &  -0.414 +- 0.016 \\
        NGC2916 &   Sb & 10.753 &  6.751 &  2.909 &  -0.599 +- 0.063 &  -0.203 +- 0.021 \\
        NGC2918 &    E & 11.443 &  5.751 &  0.647 &  -0.922 +- 0.032 &  -0.225 +- 0.008 \\
        NGC3106 &   Sa & 11.212 &  9.195 &  1.529 &  -0.726 +- 0.071 &  -0.262 +- 0.027 \\
        NGC3300 &   S0 & 10.762 &  2.903 &  2.042 &  -1.057 +- 0.078 &  -0.268 +- 0.017 \\
        NGC3381 &   Sd &  9.681 &  2.761 &  3.884 &   0.172 +- 0.196 &   0.045 +- 0.070 \\
        NGC3687 &   Sb & 10.275 &  2.949 &  2.702 &  -0.221 +- 0.130 &  -0.099 +- 0.055 \\
        NGC3811 &   Sb & 10.423 &  4.550 &  2.034 &  -0.471 +- 0.011 &  -0.210 +- 0.005 \\
        NGC3815 &   Sb & 10.353 &  3.666 &  1.539 &  -0.751 +- 0.076 &  -0.285 +- 0.031 \\
        NGC3994 &   Sb & 10.423 &  2.033 &  1.818 &  -1.249 +- 0.234 &  -0.484 +- 0.066 \\
        NGC4003 &   S0 & 11.073 &  6.482 &  2.489 &  -0.517 +- 0.082 &  -0.138 +- 0.022 \\
        NGC4149 &   Sa & 10.362 &  3.970 &  3.165 &  -0.666 +- 0.041 &  -0.320 +- 0.021 \\
        NGC4185 &   Sb & 10.671 &  8.309 &  3.982 &  -0.405 +- 0.101 &  -0.140 +- 0.030 \\
        NGC4210 &   Sb & 10.286 &  4.025 &  4.222 &  -0.160 +- 0.082 &  -0.070 +- 0.028 \\
        NGC4644 &   Sb & 10.451 &  4.329 &  2.444 &  -0.436 +- 0.140 &  -0.160 +- 0.049 \\
        NGC4711 &   Sb & 10.312 &  5.029 &  2.324 &  -0.549 +- 0.077 &  -0.244 +- 0.036 \\
       NGC4841A &    E & 11.546 &  9.665 &  0.295 &  -0.755 +- 0.278 &  -0.172 +- 0.058 \\
        NGC4956 &    E & 10.986 &  3.134 &  1.869 &  -0.845 +- 0.074 &  -0.195 +- 0.014 \\
        NGC4961 &   Sc &  9.680 &  2.671 &  0.981 &  -0.315 +- 0.093 &  -0.280 +- 0.089 \\
        NGC5000 &   Sb & 10.730 &  6.292 &  3.655 &  -0.001 +- 0.062 &  -0.022 +- 0.013 \\
        NGC5016 &   Sb & 10.235 &  3.227 &  2.176 &  -0.100 +- 0.053 &  -0.040 +- 0.025 \\
        NGC5205 &   Sb &  9.864 &  2.331 &  2.196 &  -0.557 +- 0.038 &  -0.309 +- 0.021 \\
        NGC5216 &    E & 10.505 &  4.111 &  1.116 &  -1.125 +- 0.191 &  -0.476 +- 0.093 \\
        NGC5218 &   Sa & 10.652 &  3.734 &  1.419 &   0.283 +- 0.035 &   0.082 +- 0.009 \\
        NGC5378 &   Sb & 10.583 &  5.184 &  3.227 &  -0.574 +- 0.041 &  -0.284 +- 0.022 \\
        NGC5406 &   Sb & 11.273 &  7.550 &  3.142 &  -1.171 +- 0.023 &  -0.297 +- 0.005 \\
        NGC5480 &   Sc & 10.141 &  3.374 &  1.937 &  -0.221 +- 0.089 &  -0.142 +- 0.055 \\
        NGC5485 &    E & 11.024 &  4.196 &  0.577 &  -0.819 +- 0.019 &  -0.249 +- 0.006 \\
 
\hline
\end{tabular}

\end{table*}


\begin{table*}
\contcaption{All positive galaxies}
\label{table:pars}
\begin{tabular}{@{}lcccccc}

\hline
Galaxy & G Type & log(Mstar/M$_{\sun}$) & $R_{\mathrm{e}}$/kpc & $<V_{rot}$/$\sigma>$ & Z gradient +- err & log(Z) gradient +- err \\
\hline
\hline

        NGC5520 &   Sb &  9.865 &  1.596 &  1.735 &  -1.089 +- 0.123 &  -0.437 +- 0.038 \\
        NGC5614 &   Sa & 11.298 &  5.012 &  0.506 &  -1.171 +- 0.123 &  -0.298 +- 0.025 \\
        NGC5630 &   Sd &  9.670 &  4.153 &  1.582 &   0.198 +- 0.093 &   0.220 +- 0.109 \\
        NGC5631 &   S0 & 10.928 &  2.608 & -0.133 &  -1.355 +- 0.006 &  -0.464 +- 0.002 \\
        NGC5633 &   Sb & 10.260 &  2.232 &  1.783 &  -0.147 +- 0.014 &  -0.074 +- 0.006 \\
        NGC5657 &   Sb & 10.284 &  2.904 &  1.748 &   0.147 +- 0.050 &   0.065 +- 0.021 \\
        NGC5682 &   Sc &  9.403 &  4.235 &  2.149 &  -0.077 +- 0.048 &  -0.148 +- 0.125 \\
        NGC5720 &   Sb & 10.848 &  8.969 &  3.468 &  -0.723 +- 0.032 &  -0.265 +- 0.012 \\
        NGC5732 &   Sb &  9.928 &  3.721 &  3.115 &  -0.299 +- 0.076 &  -0.166 +- 0.051 \\
        NGC5784 &   S0 & 11.216 &  5.067 &  0.903 &  -1.183 +- 0.025 &  -0.308 +- 0.006 \\
        NGC5797 &    E & 10.846 &  5.142 &  2.160 &  -1.357 +- 0.015 &  -0.475 +- 0.008 \\
        NGC5876 &   S0 & 10.901 &  2.778 &  0.814 &   0.048 +- 0.012 &   0.016 +- 0.003 \\
        NGC5888 &   Sb & 11.206 &  9.803 &  2.713 &  -0.234 +- 0.030 &  -0.066 +- 0.008 \\
        NGC5908 &   Sa & 11.223 &  7.989 &  1.857 &  -1.012 +- 0.034 &  -0.290 +- 0.009 \\
        NGC5930 &   Sa & 10.633 &  3.022 &  2.644 &  -0.346 +- 0.044 &  -0.112 +- 0.012 \\
        NGC5953 &   Sa & 10.478 &  1.411 & -0.812 &   0.558 +- 0.028 &   1.247 +- 0.053 \\
        NGC5966 &    E & 11.009 &  5.820 &  0.596 &  -0.792 +- 0.031 &  -0.266 +- 0.009 \\
        NGC5971 &   Sb & 10.317 &  2.965 &  1.401 &  -0.217 +- 0.085 &  -0.117 +- 0.045 \\
        NGC5980 &   Sb & 10.720 &  4.931 &  2.326 &  -0.379 +- 0.055 &  -0.139 +- 0.020 \\
        NGC5987 &   Sa & 11.210 &  6.907 &  1.321 &  -1.219 +- 0.019 &  -0.372 +- 0.007 \\
        NGC6004 &   Sb & 10.687 &  5.888 &  4.731 &  -0.679 +- 0.030 &  -0.253 +- 0.012 \\
        NGC6020 &    E & 11.001 &  5.679 &  0.885 &  -1.008 +- 0.116 &  -0.335 +- 0.036 \\
        NGC6021 &    E & 11.006 &  3.126 &  1.405 &  -1.546 +- 0.283 &  -0.465 +- 0.071 \\
        NGC6032 &   Sb & 10.527 &  8.220 & -6.494 &   0.450 +- 0.032 &   0.135 +- 0.011 \\
        NGC6060 &   Sb & 10.934 &  8.918 &  2.830 &  -1.837 +- 0.071 &  -0.691 +- 0.019 \\
        NGC6063 &   Sb & 10.140 &  4.076 &  3.530 &   0.317 +- 0.051 &   0.168 +- 0.027 \\
        NGC6081 &   S0 & 11.118 &  4.351 &  1.141 &  -0.958 +- 0.024 &  -0.277 +- 0.007 \\
        NGC6125 &    E & 11.384 &  6.978 &  0.303 &  -1.302 +- 0.168 &  -0.344 +- 0.035 \\
        NGC6132 &   Sb & 10.212 &  5.055 &  1.775 &  -0.487 +- 0.069 &  -0.195 +- 0.030 \\
        NGC6146 &    E & 11.629 &  9.123 &  1.018 &  -1.083 +- 0.090 &  -0.318 +- 0.024 \\
        NGC6150 &    E & 11.426 &  7.211 &  1.228 &  -0.164 +- 0.022 &  -0.036 +- 0.005 \\
        NGC6168 &   Sc &  9.863 &  4.807 &  1.529 &   0.335 +- 0.180 &   0.158 +- 0.119 \\
        NGC6173 &    E & 11.725 & 23.189 &  0.374 &  -0.744 +- 0.106 &  -0.204 +- 0.031 \\
        NGC6186 &   Sb & 10.569 &  4.121 &  2.314 &  -0.066 +- 0.101 &  -0.023 +- 0.034 \\
        NGC6278 &   S0 & 10.919 &  2.148 &  1.097 &  -0.039 +- 0.033 &  -0.009 +- 0.008 \\
        NGC6310 &   Sb & 10.561 &  5.610 &  2.950 &  -0.407 +- 0.038 &  -0.150 +- 0.014 \\
        NGC6314 &   Sa & 11.211 &  5.815 &  0.832 &  -1.305 +- 0.023 &  -0.419 +- 0.003 \\
        NGC6394 &   Sb & 10.896 &  8.647 &  2.613 &   0.318 +- 0.046 &   0.118 +- 0.016 \\
        NGC6427 &   S0 & 10.751 &  1.872 &  0.805 &  -1.056 +- 0.120 &  -0.221 +- 0.025 \\
        NGC6478 &   Sc & 11.014 & 10.997 &  2.716 &  -1.021 +- 0.057 &  -0.359 +- 0.019 \\
        NGC6497 &   Sa & 11.037 &  2.848 &  1.955 &  -0.581 +- 0.046 &  -0.152 +- 0.012 \\
        NGC6515 &    E & 11.193 &  9.017 &  0.098 &  -0.337 +- 0.292 &  -0.087 +- 0.074 \\
        NGC6762 &   Sa & 10.383 &  1.927 &  1.687 &  -0.793 +- 0.191 &  -0.234 +- 0.062 \\
        NGC6941 &   Sb & 10.943 &  9.044 &  3.257 &  -0.785 +- 0.053 &  -0.225 +- 0.014 \\
        NGC6945 &   S0 & 11.389 &  3.417 &  1.703 &  -1.473 +- 0.024 &  -0.344 +- 0.005 \\
        NGC7025 &   S0 & 11.527 &  4.633 &  0.963 &  -1.010 +- 0.054 &  -0.241 +- 0.013 \\
        NGC7047 &   Sb & 10.791 &  7.471 &  3.463 &  -0.720 +- 0.066 &  -0.225 +- 0.020 \\
        NGC7194 &    E & 11.445 & 10.057 &  1.201 &  -0.885 +- 0.013 &  -0.236 +- 0.003 \\
        NGC7311 &   Sa & 11.069 &  3.957 &  1.959 &  -0.869 +- 0.037 &  -0.209 +- 0.008 \\
        NGC7321 &   Sb & 10.931 &  7.425 &  3.131 &  -0.729 +- 0.037 &  -0.212 +- 0.011 \\
 
\hline
\end{tabular}

\end{table*}


\begin{table*}
\contcaption{A table continued from the previous one.}
\label{table:continued}
\begin{tabular}{@{}lcccccc}

\hline
Galaxy & G Type & log(Mstar/M$_{\sun}$) & $R_{\mathrm{e}}$/kpc & $<V_{rot}$/$\sigma>$ & Z gradient +- err & log(Z) gradient +- err \\
\hline
\hline

        NGC7364 &   Sa & 10.882 &  4.261 &  1.912 &  -0.507 +- 0.053 &  -0.174 +- 0.017 \\
       NGC7436B &    E & 11.914 & 14.159 &  0.247 &  -1.397 +- 0.289 &  -0.327 +- 0.073 \\
        NGC7466 &   Sb & 10.748 &  7.006 &  2.024 &  -0.775 +- 0.113 &  -0.283 +- 0.037 \\
        NGC7489 &   Sb & 10.501 &  8.720 &  2.334 &  -0.533 +- 0.107 &  -0.354 +- 0.079 \\
        NGC7549 &   Sb & 10.599 &  6.597 &  1.939 &  -0.690 +- 0.044 &  -0.395 +- 0.029 \\
        NGC7562 &    E & 11.247 &  5.211 &  0.222 &  -0.371 +- 0.053 &  -0.081 +- 0.012 \\
        NGC7563 &   Sa & 10.963 &  2.949 &  0.750 &  -0.633 +- 0.031 &  -0.131 +- 0.006 \\
        NGC7591 &   Sb & 10.760 &  5.571 &  2.449 &  -0.061 +- 0.031 &  -0.024 +- 0.015 \\
        NGC7608 &   Sb & 10.094 &  4.922 &  2.738 &   0.220 +- 0.128 &   0.130 +- 0.075 \\
        NGC7619 &    E & 10.944 &  9.350 &  0.273 &  -0.771 +- 0.035 &  -0.194 +- 0.008 \\
        NGC7623 &   S0 & 10.981 &  2.728 &  0.954 &  -0.974 +- 0.024 &  -0.247 +- 0.006 \\
        NGC7625 &   Sa & 10.125 &  1.604 &  3.727 &  -0.304 +- 0.105 &  -0.208 +- 0.068 \\
        NGC7631 &   Sb & 10.529 &  4.530 &  2.578 &  -0.537 +- 0.092 &  -0.206 +- 0.037 \\
        NGC7653 &   Sb & 10.499 &  3.748 &  1.878 &  -0.873 +- 0.049 &  -0.299 +- 0.018 \\
        NGC7671 &   S0 & 10.956 &  3.101 &  1.373 &  -1.026 +- 0.035 &  -0.220 +- 0.006 \\
        NGC7683 &   S0 & 11.019 &  3.668 &  1.308 &  -0.656 +- 0.038 &  -0.161 +- 0.009 \\
        NGC7684 &   S0 & 10.986 &  3.544 &  1.642 &  -1.442 +- 0.106 &  -0.340 +- 0.020 \\
        NGC7711 &    E & 11.053 &  4.230 &  1.275 &  -1.290 +- 0.079 &  -0.383 +- 0.021 \\
        NGC7716 &   Sb & 10.389 &  3.794 &  2.087 &  -0.710 +- 0.019 &  -0.299 +- 0.010 \\
        NGC7722 &   Sa & 11.245 &  5.943 &  1.547 &  -1.304 +- 0.025 &  -0.384 +- 0.006 \\
        NGC7738 &   Sb & 11.079 &  6.951 &  1.712 &   0.940 +- 0.038 &   0.214 +- 0.031 \\
        NGC7787 &   Sa & 10.621 &  5.298 &  1.588 &   0.096 +- 0.186 &   0.068 +- 0.084 \\
        NGC7800 &   Sd &  9.289 &  3.961 &  1.027 &   0.296 +- 0.057 &   0.383 +- 0.064 \\
        NGC7819 &   Sc & 10.389 &  8.227 &  2.123 &   0.082 +- 0.020 &   0.024 +- 0.011 \\
       UGC00005 &   Sb & 10.831 &  8.405 &  3.256 &  -0.081 +- 0.038 &  -0.026 +- 0.013 \\
       UGC00036 &   Sa & 11.002 &  4.480 &  1.822 &   0.077 +- 0.094 &   0.021 +- 0.025 \\
       UGC00148 &   Sc & 10.112 &  6.089 &  1.539 &   0.207 +- 0.134 &   0.153 +- 0.096 \\
       UGC00312 &   Sd &  9.780 &  6.198 &  1.082 &   0.136 +- 0.055 &   0.249 +- 0.118 \\
  UGC00335NED02 &    E & 10.783 &  6.906 &  0.559 &  -1.075 +- 0.095 &  -0.524 +- 0.046 \\
       UGC00809 &   Sc &  9.686 &  5.875 &  1.766 &  -0.599 +- 0.113 &  -0.390 +- 0.089 \\
       UGC00841 &   Sb & 10.011 &  6.585 &  3.243 &  -0.622 +- 0.181 &  -0.474 +- 0.146 \\
       UGC00987 &   Sa & 10.614 &  4.085 &  2.344 &  -0.265 +- 0.080 &  -0.092 +- 0.027 \\
       UGC01057 &   Sc & 10.105 &  6.263 &  1.915 &  -0.321 +- 0.115 &  -0.132 +- 0.047 \\
       UGC01271 &   S0 & 10.827 &  3.317 &  1.297 &  -0.828 +- 0.041 &  -0.176 +- 0.009 \\
       UGC02222 &   S0 & 10.743 &  3.689 &  1.705 &  -1.240 +- 0.189 &  -0.347 +- 0.047 \\
       UGC02229 &   S0 & 10.890 &  9.629 &  1.454 &  -1.681 +- 0.049 &  -0.604 +- 0.019 \\
       UGC03151 &   Sa & 10.757 &  6.122 &  2.369 &  -0.862 +- 0.718 &  -0.421 +- 0.317 \\
       UGC03944 &   Sb &  9.997 &  4.712 &  1.363 &  -0.343 +- 0.071 &  -0.211 +- 0.042 \\
       UGC03969 &   Sb & 10.679 &  8.591 &  2.547 &   0.049 +- 0.134 &   0.026 +- 0.062 \\
       UGC03995 &   Sb & 10.922 &  8.514 &  2.443 &  -0.516 +- 0.026 &  -0.179 +- 0.012 \\
       UGC04029 &   Sc & 10.330 &  8.052 &  2.859 &  -0.661 +- 0.043 &  -0.280 +- 0.017 \\
       UGC04132 &   Sb & 10.765 &  8.285 &  2.580 &  -0.513 +- 0.100 &  -0.214 +- 0.039 \\
       UGC04145 &   Sa & 10.959 &  3.061 &  1.343 &  -0.062 +- 0.113 &  -0.029 +- 0.052 \\
       UGC04197 &   Sa & 10.712 &  5.773 &  2.143 &  -0.259 +- 0.111 &  -0.092 +- 0.040 \\
       UGC04280 &   Sb & 10.137 &  2.696 &  1.759 &   0.216 +- 0.068 &   0.100 +- 0.031 \\
       UGC05108 &   Sb & 10.889 &  5.350 &  1.297 &   0.430 +- 0.157 &   0.139 +- 0.052 \\
       UGC05113 &   S0 & 11.101 &  3.927 &  1.595 &  -1.776 +- 0.159 &  -0.443 +- 0.034 \\
  UGC05498NED01 &   Sa & 10.805 &  6.052 &  1.922 &  -0.582 +- 0.062 &  -0.186 +- 0.020 \\
       UGC05598 &   Sb & 10.232 &  6.169 &  1.514 &  -0.213 +- 0.108 &  -0.125 +- 0.065 \\
       UGC05771 &    E & 11.317 &  6.512 &  1.511 &  -0.944 +- 0.030 &  -0.241 +- 0.007 \\
       UGC06036 &   Sa & 11.172 &  5.386 &  1.365 &  -0.341 +- 0.075 &  -0.079 +- 0.016 \\
 
\hline
\end{tabular}

\end{table*}


\begin{table*}
\contcaption{All positive galaxies}
\label{table:pars}
\begin{tabular}{@{}lcccccc}

\hline
Galaxy & G Type & log(Mstar/M$_{\sun}$) & $R_{\mathrm{e}}$/kpc & $<V_{rot}$/$\sigma>$ & Z gradient +- err & log(Z) gradient +- err \\
\hline
\hline

       UGC06312 &   Sa & 11.031 &  6.079 &  1.437 &  -1.062 +- 0.028 &  -0.338 +- 0.010 \\
       UGC07145 &   Sb & 10.355 &  7.652 &  4.782 &  -0.383 +- 0.079 &  -0.151 +- 0.036 \\
       UGC08107 &   Sa & 11.066 &  9.562 &  0.917 &  -0.487 +- 0.042 &  -0.183 +- 0.016 \\
       UGC08231 &   Sd &  9.133 &  3.258 &  2.383 &   0.128 +- 0.096 &   0.275 +- 0.204 \\
       UGC08234 &   S0 & 11.135 &  4.669 &  1.725 &  -0.881 +- 0.062 &  -0.224 +- 0.016 \\
       UGC08778 &   Sb & 10.245 &  3.470 &  1.925 &   0.192 +- 0.051 &   0.087 +- 0.023 \\
       UGC08781 &   Sb & 11.056 &  8.109 &  2.485 &  -0.541 +- 0.039 &  -0.162 +- 0.011 \\
       UGC09067 &   Sb & 10.582 &  7.962 &  2.888 &  -0.633 +- 0.041 &  -0.223 +- 0.015 \\
       UGC09476 &   Sb & 10.207 &  4.935 &  1.977 &  -0.015 +- 0.066 &  -0.004 +- 0.032 \\
       UGC09537 &   Sb & 11.220 & 12.350 &  1.769 &  -0.621 +- 0.123 &  -0.209 +- 0.041 \\
       UGC09542 &   Sc & 10.316 &  8.124 &  2.653 &   0.091 +- 0.031 &   0.042 +- 0.016 \\
       UGC09665 &   Sb &  9.996 &  3.296 &  2.199 &   0.170 +- 0.122 &   0.117 +- 0.081 \\
       UGC09873 &   Sb & 10.098 &  8.342 &  2.534 &  -0.184 +- 0.038 &  -0.119 +- 0.027 \\
       UGC09892 &   Sb & 10.297 &  6.537 &  3.838 &  -0.800 +- 0.159 &  -0.387 +- 0.076 \\
       UGC10097 &    E & 11.458 &  6.049 &  0.804 &  -0.918 +- 0.080 &  -0.233 +- 0.018 \\
       UGC10123 &   Sa & 10.521 &  4.752 &  1.825 &   0.035 +- 0.255 &   0.012 +- 0.084 \\
       UGC10205 &   S0 & 10.997 &  8.835 &  1.012 &  -0.812 +- 0.079 &  -0.325 +- 0.031 \\
       UGC10297 &   Sc &  9.460 &  2.966 &  0.760 &  -0.214 +- 0.072 &  -0.208 +- 0.092 \\
       UGC10331 &   Sc &  9.887 &  5.995 & -0.080 &   0.360 +- 0.082 &   0.390 +- 0.087 \\
       UGC10337 &   Sb & 11.033 & 10.586 &  3.189 &  -0.779 +- 0.106 &  -0.272 +- 0.034 \\
       UGC10380 &   Sb & 11.009 &  7.447 &  1.179 &  -0.473 +- 0.200 &  -0.187 +- 0.075 \\
       UGC10384 &   Sb & 10.271 &  4.072 &  1.788 &   0.305 +- 0.048 &   0.175 +- 0.029 \\
       UGC10388 &   Sa & 10.817 &  3.545 &  1.839 &  -0.125 +- 0.031 &  -0.048 +- 0.012 \\
       UGC10693 &    E & 11.507 & 13.353 &  0.291 &  -1.643 +- 0.114 &  -0.529 +- 0.042 \\
       UGC10695 &    E & 11.300 & 14.274 &  8.761 &  -0.380 +- 0.126 &  -0.225 +- 0.014 \\
       UGC10710 &   Sb & 10.986 & 11.724 &  1.953 &  -0.789 +- 0.105 &  -0.285 +- 0.040 \\
       UGC10796 &   Sc &  9.450 &  4.372 &  0.691 &  -0.402 +- 0.118 &  -0.289 +- 0.105 \\
       UGC10811 &   Sb & 10.874 &  7.654 &  1.815 &  -0.395 +- 0.055 &  -0.130 +- 0.018 \\
       UGC10905 &   S0 & 11.607 &  8.502 &  1.714 &  -1.359 +- 0.104 &  -0.384 +- 0.031 \\
       UGC10972 &   Sb & 10.425 &  8.038 &  4.139 &  -0.405 +- 0.052 &  -0.175 +- 0.026 \\
       UGC11228 &   S0 & 11.093 &  4.933 &  1.529 &  -0.831 +- 0.037 &  -0.195 +- 0.011 \\
       UGC11649 &   Sa & 10.568 &  4.995 &  3.611 &  -0.699 +- 0.026 &  -0.264 +- 0.010 \\
       UGC11717 &   Sa & 10.842 &  7.832 &  2.269 &  -0.685 +- 0.046 &  -0.141 +- 0.008 \\
       UGC12054 &   Sc &  8.993 &  2.181 & -1.959 &   0.033 +- 0.128 &   0.039 +- 0.147 \\
       UGC12127 &    E & 11.369 & 20.769 & -0.310 &  -1.321 +- 0.587 &  -0.428 +- 0.226 \\
       UGC12185 &   Sb & 10.670 &  5.653 &  1.915 &  -0.105 +- 0.066 &  -0.031 +- 0.019 \\
       UGC12274 &   Sa & 11.152 &  9.026 &  2.010 &  -0.534 +- 0.069 &  -0.165 +- 0.021 \\
       UGC12494 &   Sd &  9.445 &  6.107 &  0.714 &  -0.093 +- 0.318 &  -0.060 +- 0.227 \\
       UGC12518 &   Sb & 10.255 &  3.287 &  2.998 &  -0.400 +- 0.098 &  -0.155 +- 0.041 \\
       UGC12519 &   Sc & 10.039 &  6.486 &  3.846 &  -0.062 +- 0.153 &  -0.016 +- 0.039 \\
       UGC12810 &   Sb & 10.735 & 11.530 &  2.740 &  -0.181 +- 0.045 &  -0.066 +- 0.014 \\
       UGC12816 &   Sc &  9.823 &  5.995 &  2.018 &  -0.134 +- 0.059 &  -0.072 +- 0.030 \\
       UGC12857 &   Sb &  9.747 &  3.305 &  0.785 &  -0.226 +- 0.741 &  -0.091 +- 0.354 \\
     VV488NED02 &   Sb & 10.365 &  7.979 &  2.418 &   0.400 +- 0.167 &   0.118 +- 0.058 \\

\hline
\end{tabular}

\end{table*}

\bsp

\label{lastpage}

\end{document}